\documentclass[a4paper,12pt]{article}
\usepackage{jcappub} % for details on the use of the package, please
\usepackage{cite}
\usepackage{graphicx}
\usepackage{enumerate}
\usepackage{hyperref}

\newcommand{\mnras}{Mon. Not. R. Astron. Soc.}

\newcommand{\prd}{Phys. Rev. D}
\newcommand{\apj}{Astrophys. J.}

\newcommand{\apjl}{Astrophys. J. Lett.}

\usepackage{cancel}
\usepackage{color}
\usepackage{graphicx}
\usepackage{amsmath}
\usepackage{amsfonts}
\usepackage{amssymb}
\usepackage{rotating}
\usepackage{booktabs}
\usepackage{xcolor}
\usepackage{soul}

\def\comment#1{}

\usepackage{graphicx}

\title{\boldmath 
Trapped fireshell (halo) of photons and pairs around black-hole horizon: source for ultra-high-energy particles}

\author{She-Sheng Xue}

\affiliation{ICRANet Piazzale della Repubblica, 10 -65122, Pescara, Italy
\\ Physics Department, Sapienza University of Rome, 
Rome, Italy\\INFN, Sezione di Perugia, %Via A. Pascoli, I-06123, 
Perugia, Italy
\\ICTP-AP, University of Chinese Academy of Sciences, Beijing, China
}

\emailAdd{xue@icra.it, she-sheng.xue@cern.ch} 

\abstract{We study the Compton-rocket effect of 
strong radiation force accelerating electrons in an opaque fireshell (or fire spot) of dense photons and 
electron-positron pairs, whose temperature is spatially inhomogeneous and exceeds the electron mass.
We find the possibility of the charged-particle acceleration and the avalanche runaway process, leading to a 
non-trivial probability of ultra-high-energy (UHE) electrons and protons, which subsequently produce very-high-energy (VHE) photons and neutrinos. 
In a simplified one-dimensional model, we qualitatively show such peculiar dynamics using the fireball, Gamma-Ray Burst central engine, whose inner part inflows and forms a gravitationally trapped fireshell (halo) around the horizon of a black hole. 
The fireshell is metastable, cooling via UHE particle emissions and blackbody radiation. 
We calculate the UHE particle luminosity 
varying in time, and discuss the peculiar features of such produced UHE particles, which lead to VHE particles, in connection with possible numerical simulations, observations and experiments.
}

\begin{document}

\maketitle
\flushbottom

\newpage

\section{Introduction}

The discovery of ultra-high-energy (UHE $\sim 10^{18-20}$ eV) charged particles (cosmic rays) has been  
for decades. The acceleration mechanism of UHE charged particles remains mysterious, although much theoretical and observational progress has been made to understand their nature and identify their sources (see review \cite{Globus2025a}). Very-high-energy (VHE $\sim 10^{12-15}$ eV) neutral particles (photons and neutrinos) have been recently observed by the LHAASO \cite{LHAASO:2021gok}, MAGIC \cite{Acciari2019}, and IceCube \cite{IceCube:2023ame} collaborations. Much theoretical attention has been given to understanding the physical origin 
of UHE charged particles and VHE neutral particles, and identifying astrophysical sources where they come from.

It is physically reasonable to assume that these highly energetic particles may originate from compact and powerful astrophysical sources, such as accretion disks, massive core collapses, binary coalescence and active galactic nuclei. 
In these sources, there may exist {\it highly energetic arenas}, where strong gravitational fields, dense radiation fields, and charged particles interact non-linearly with one another. 
These interactions possibly result in the 
high-energy processes of accelerating charged particles, leading to UHE and VHE particles.   

One of these powerful astrophysical sources can be the central engine for Gamma-ray bursts (GRBs), an opaque fireballs of large ultra-relativistic particle (photon and electron-positron pair) energy density and pressure \cite{Rees:1992ek}, and see reviews \cite{2004RvMP...76.1143P,2006RPPh...69.2259M,2014ARA&A..52...43B,2015JHEAp...7...73D,2015PhR...561....1K,zhang_2018, Ruffini2010,2012grb..book.....K}. 
Within the general relativity (GR) framework, we studied \cite{ Ruffini1999, Ruffini2000} the hydrodynamical outflow of fireballs by using also the Boltzmann rate equation for electrons and positrons (pairs) annihilation to photons, obtaining the relation between the photon heat and baryon kinetic energies for main bursts and afterglow radiations. 
We will first use the fireball, Gamma-Ray Bursts (GRBs) central engine,
to show a possible {\it highly energetic arena}, a 
fireshell or fire spot, then illustrate our studies and results of the high-energy interactions and processes that produce UHE and VHE particles in such an arena.

We will study in Section \ref{trapp} that through the gravitational force and thermal pressure, the fireball's inner part (fireshell, halo around a black hole horizon) is trapped and forms a metastable thermal fluid of photons and pairs at high energy density and temperature. Using such an opaque horizon halo, we study the Compton-rocket effect on accelerating charged particles (Section \ref{Compton}) and the avalanche runaway process (Section \ref{run}) leading to UHE electrons and protons. 
We calculate the luminosity of UHE particles and the halo cooling in Section \ref{lumep}, and discuss the peculiar features of UHE and VHE particles produced in this dynamics, in connection with numerical simulations, observations, and experiments in the final Section \ref{obs}. We use the $\hbar=c=1$ and Compton unit of the electron mass $m_e$, unless otherwise specified in necessary cases for clarification.

\begin{figure}
\centering
\includegraphics[width=5.0in]{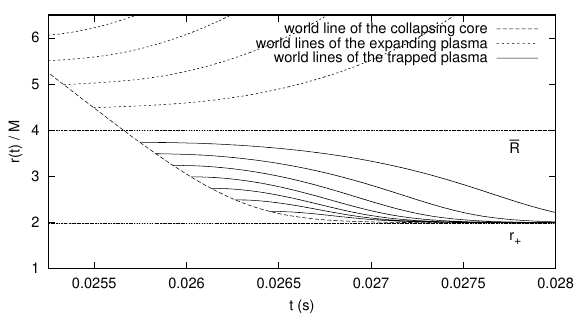}
\caption{Figure 2 of Ref.~\cite{Ruffini2003} shows the world lines of fireball hydrodynamics expanding outflow (dot line) and trapped 
inflow (solid line) in the Schwarzschild geometry of a collapsing massive stellar core (dashed line), which forms a black hole of mass $M$. Near the horizon radius $r_+=2M$ of a formed black hole, the gravitational time dilation $\delta t\gg \delta\tau$ for an observer at a distance is evident; $\delta\tau$ is a finite proper time (distance) for a local observer.
The separatrix radius between out and in flows is $\bar R = 4M$, and the trapped inflow energy is comparable with %($0.53$) 
the expanding outflow energy. The total inner and outer fireball energy gain from the collapsing system's gravitational binding energy, and it should 
be much smaller than $-(M^2/r_\infty-M^2/r_+)/2=M/4$ 
in the Newtonian approximation. $G=1$ in this figure.
}\label{fallin}
\end{figure}

\section{Trapped fireshell (halo) around black hole horizon}\label{trapp}

\subsection{Trapped fireshell (halo) of energetic and dense photons and pairs}

Although the dynamics of fireball formation 
are still under investigation \footnote{In Refs.~\cite{Xue_2021, Xue2025}, we present the detailed studies of the possibility of how fireballs or jets form during gravitational collapses of massive rotating stellar cores or binary coalescence: heavy hadrons gain kinetic energy from gravitational potential,
and their heavy-ion collisions produce photons and pairs, resulting in the fireball or jet formation.}, the hydrodynamical outflow of fireballs accounting for GRBs has been well understood upon the assumptions that entire fireballs are propelled out in a flat spacetime. Nevertheless, a scrutiny GR study \cite{Ruffini2003} 
investigates the hydrodynamical evolution of fireballs in the Schwarzschild geometry of a massive stellar core, which undergoes a gravitational collapse process to form a black hole of mass $M$.
The result shows in Fig.~\ref{fallin} that the fireball's inner part $(r<\bar R)$ is trapped due to strong gravitational attraction, prevailing over thermal pressure repulsion. On the contrary, its outer part $(r>\bar R)$, where gravitational attraction is weak and thermal pressure repulsion prevails, 
hydrodynamically outflows, undergoing internal and external shocks in its evolution, accounting for the GRB phenomenon as usually described. 
The radius separatix between the inner and outer parts is $\bar R\approx 4GM=2r_+$ near the black hole horizon $r_+=2GM$ \footnote{ 
We can approximately estimate the separatrix $\bar R$ based on the balance between thermal repulsive pressure and gravitational attractive force
$4\pi r_+^2\rho_\gamma(r_+)c \sim G m_\gamma M/r_+^2$ on the fireball inner part of energy mass $m_\gamma\sim \rho_\gamma(r_+)4\pi r_+^2\bar R$ and fireball
density $\rho_\gamma(r_+)$ at the horizon.}. 

The gravitational collapsing system comprises a cold 
baryon core and a hot fireball of relativistic particles. 
When the spacetime horizon is approaching, the final phase of the gravitational collapse of the nonrelativistic baryon matter core and the relativistic particles' fluid 
(fireball inner part), which becomes denser, 
must be nontrivial. It can neither be a free-fall process of the time scale $\sim GM/c$, nor a rapid transit process, like a GRB prompt emission of an expanding fireball at its transparency. In the final 
phase of gravitational collapses, the gravitational time dilation effect is significant for an observer at a distance, see Fig.~\ref{fallin} and its caption.

The trapped fireball's inner part, we call the fireshell (or halo) for short, is highly energetic and dense, comprising numerous ultra-relativistic particles and antiparticles, which are represented by 
photons $\gamma$, electrons $e^-$ and positrons $e^+$ pairs. They have zero chemical potential and are electrically neutral in total. It is completely opaque, undergoing the rapid back-and-forth process 
\begin{equation}
\gamma+\gamma \leftrightarrow e^++e^-,
\label{bandf}
\end{equation}
which establishes a local thermal equilibrium at high temperatures and densities. Since 
the spacetime length scales of the back-and-forth process (\ref{bandf}) are very small, we describe all photons $\gamma$ and pairs $e^+e^-$ together as a radiation fluid, rather than identifying them as individual photons, electrons $e^-$ and positrons $e^+$ \footnote{In other words, relativistic fermions always pair with their antiparticles via exchanging two photons. Therefore, we 
ignore the notion of anti-particles like positrons $e^+$ in this article.}. 
We treat a local element of the trapped fireball as a perfect radiation fluid of large energy density $\rho_\gamma(r)$ and pressure $P_\gamma(r)$  
with the Equation of State 
$P_\gamma(r)=\rho_\gamma (r)/3$. 
They are inhomogeneous as functions of radius $r$, due to the presence of the gravitational field of a massive collapsing stellar core or black hole. Once more, we stress that the total thermal energy of the trapped fireball's inner part is comparable to that of the expanding fireball's outer part 
for GRBs \cite{Ruffini2003}. 

Moreover, in distinct contrast with the
gravitational collapse processes of pressure-less baryon dust, the dynamics of inflow contraction of the opaque fireshell 
is much more complex, because the gravitational potential, the fireshell density and pressure increase. 
To study such an inflow contraction in time, we should, in principle, consider the inhomogeneous radiative fluid and 
Einstein equation in the spherically symmetric Lema\^itre spacetime \cite{Lemaitre:1933gd} \footnote{The Lema\^itre metric reduces to the Lema\^itre-Tolman-Bondi one in the pressureless dust case, whose solutions 
are well known. However, despite many years of work, the solution of
the Lema\^itre metric with a general perfect fluid source 
is not known. %, and to the Friedmann-Lema\^itre-Robertson-Walker metric in a homogenous presure $P=P(t)$.
} in the presence of a black hole, 
whose metric elements 
are functions of the coordinate radius $r$ and time $t$.  
Observe that (i) the fireshell inflow contraction should be slow in time because of its pressure being very large; 
(ii) the fireshell is deeply opaque so that its thermal energy dissipation rate 
is small. Therefore, the fireshell in Lema\^itre spacetime should be a metastable configuration undergoing slow contraction adiabatically. Namely, the Lema\^itre metric elements, the fireshell energy density and pressure should be slowly varying in time. 
Thus, we approximately adopt the static limit, all quantities being independent of time, the Lema\^itre spacetime reduces to the Tolman–Oppenheimer–Volkoff (TOV) one \cite{Lasky:2006mg} in the interior of the fireshell with the exterior in Schwarzschild coordinates.

\subsection{Tolman–Oppenheimer–Volkoff equation for halo energy density}\label{tovs}

In this section, we adopt the spherically symmetric TOV equation for the fireshell to study the gravitational equilibrium configuration of 
ultra-relativistic photons and pairs trapped in the Schwarzschild spacetime near the horizon,
\begin{eqnarray}
{\frac {dP_\gamma}{dr}}=-{\frac {Gm(r)}{r^{2}}}\rho_\gamma \left(1+{\frac {P_\gamma}{\rho_\gamma}}\right)\left(1+\frac {4\pi r^{3}P_\gamma}{m(r)}\right)\left(1-{\frac {2Gm(r)}{r}}\right)^{-1},
\label{tov00}
\end{eqnarray}
in the Schwarzschild geometry of radial coordinate $r$ and time $t$ of an observer at infinity.
The TOV equation describes the equilibrium balance between the gravitational and radiation forces. 
Within the sphere of a radius $r$, the total energy mass
${\textstyle m(r)=M+E_\gamma(r)}\approx M$ and the photon-pair fluid energy $E_\gamma(r)$ is much smaller than the black hole mass $M$. 

\begin{figure}
\centering
\includegraphics[width=3.2in,height=2.2in]{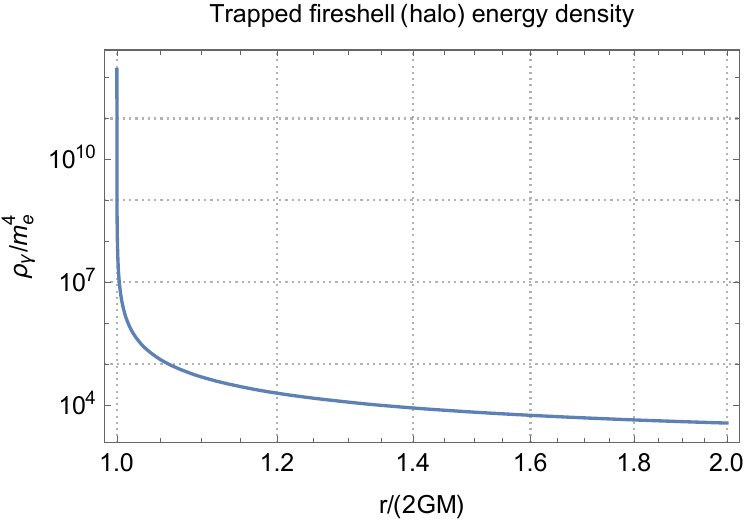}%height
\includegraphics[width=3.2in,height=2.2in]{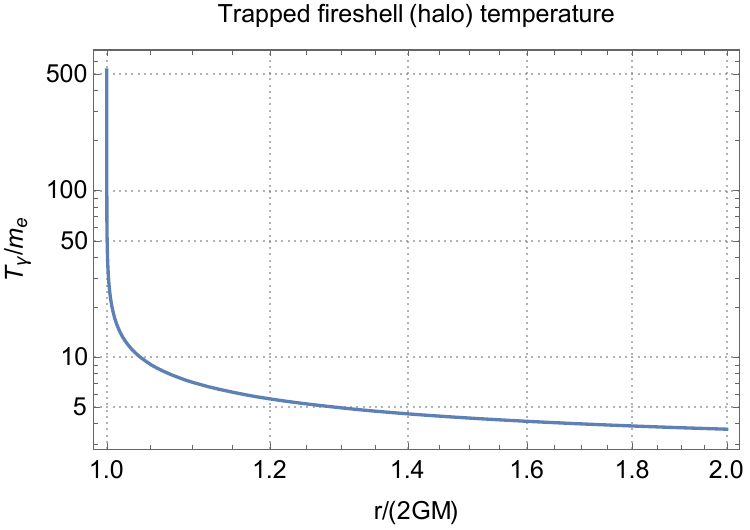}%height
\caption{Left: the trapped fireshell (halo) energy density $\rho_\gamma$, solution to (\ref{tov00}), is expressed in units of $m_e^4$ and as functions of radius $r$, which is in 
the unit of the horizon radius 
$r_+=2GM=2.33\times 10^6$ cm for the black hole mass $M=7.75 M_\odot$ (\ref{bhcon}). 
% and total trapped energy (\ref{totale}) $E_\gamma\approx 2.0\times 10^{-4}M$. 
Right: the trapped fireshell (halo) temperature 
$T_\gamma=(\rho_\gamma/\sigma_{_{\rm SB}}d_f)^{1/4}$ (\ref{tem}) is expressed in units of the electron mass $m_e\approx 0.5 {\rm MeV}$. 
The maximal temperature $T^+_\gamma$ near the horizon rapidly decreases to the asymptotic value $\approx 2m_e$ near the outer boundary $2r_+=4GM$, where the electron-positron pair annihilation takes place.
The Compton energy density $m_e^4\approx 4.54\times 10^{21}{\rm erg/cm}^3$ is of the same order as the critical energy density
$\rho_c=E_c^2/(8\pi)   %= m_e^4/(e^28\pi)=m_e^4/(\alpha 8\pi)
=5.45 m_e^4$,
where $E_c=m_e^2/e$ is the critical electric field for electron and positron pair production in the semi-classical approximation \cite{Ruffini2010}.
}\label{trapped}
\end{figure}

Rescaling the radius $r$ in the unit of $2GM$ and 
density $\rho_\gamma$ in the unit of ``average density''
$\rho_{_M}=M/V_M$, where ``volume" $V_M=(4\pi/3)(2GM)^3$, we recast the TOV equation (\ref{tov00}) as a dimensionless equation
\begin{eqnarray}
{\frac {d}{dr}}\left(\frac {\rho_\gamma}{\rho_{_M}}\right)=-\frac {3}{2r^2}\left(\frac {\rho_\gamma}{\rho_{_M}}\right)\left[1+r^3\left(\frac {\rho_\gamma}{\rho_{_M}}\right)\right]\left(1-{\frac {1}{r}}\right)^{-1},
\label{tov0}
\end{eqnarray}
for the density $\rho_\gamma$. 
We consider a massive stellar baryon core of neutron stars at the nuclear saturation density $\rho_n\approx 0.16 \,{\rm fm}^{-3}=1.6\times 10^{38}/{\rm cm}^3$, gravitationally collapsing to a compact object of size about $2GM$ of the black hole horizon. The ``average density'' $\rho_{_M}=M/V_M$ varies from $(2-10)\rho_n$ for the mass $M$ varying from $(3-10)M_{\odot}$. 
The value $\rho_{_M}=10\rho_n$ effectively 
corresponds to a black hole of the mass $M$ 
and horizon radius $r_+$  
\begin{eqnarray}
M= 7.75 M_{\odot},\quad r_+=2GM=2.33\times 10^6 {\rm cm}.
\label{bhcon}
\end{eqnarray}

The photon-pair pressure $P_\gamma(r)=\rho_\gamma (r)/3$ should be maximum near the horizon, where the gravitational force is the strongest. To illustrate the fireshell properties, we choose the photon-pair density near the horizon
\begin{eqnarray}
\rho_\gamma^+\equiv \rho_\gamma(r_+)=\eta\rho_{_M}=\rho_n,~~~\eta=0.1,
\label{bcon}
\end{eqnarray}
as the boundary conditions for the TOV equation (\ref{tov0}).
The reason for $\rho_\gamma^+\sim \rho_n$ is that at high energies, many species of relativistic particles, lepton-antilepton pairs and photons, light quark-antiquark pairs and gluons, could be excited, contributing to the energy density $\rho_\gamma^+$. It turns out to be a fluid of lepton(pair)-photon and quark(pair)-gluon plasma. For the sake of simplified notations, we use the photon ($\gamma$) and pair ($e^-e^+$) notations 
representing all possible gauge bosons and fermion-antifermion pairs, and use the subscript ``$\gamma$'' for the quantities of the trapped fireshell of photons and pairs. 

Solving the TOV equation (\ref{tov0}) with boundary conditions (\ref{bcon}), we show in Fig.~\ref{trapped} (left) that the energy density $\rho_\gamma$ of photons and pairs is maximal near the horizon $r_+$ and rapidly decreases as the radius increases to $2r_+$. 
The ``trapped fireshell (halo)'' is named for such an energetic radiation fluid configuration of photons and pairs near the black hole horizon, where the maximal gravity effect confines the opaque radiation fluid of photons and pairs at high densities. 
The macroscopic time scale of hydrodynamical relaxation for establishing a trapped fireshell 
is about $\bar R /v_s$, where the fluid sound velocity $v_s=(P_\gamma/\rho_\gamma)^{1/2}=c/\sqrt{3}$.

The total trapped fireshell energy is an integration over the 
energy density $\rho_\gamma$ from the horizon $r_+=2GM$ up to the outer boundary $\bar R= 4GM=2r_+$
\begin{eqnarray}
 %   E_\gamma(r)&=&3M \int_1^2\frac{r^2}{(1-\frac{1}{r})^{1/2}}\frac{\rho_\gamma(r)}{\rho_{_M}}dr=\chi M,\nonumber\\
     E_\gamma&=&\int_{r_+}^{2r_+}\rho_\gamma(r)\frac{4\pi r^2dr}{(1-\frac{r_+}{r})^{1/2}}=\rho_\gamma^+ 4\pi r_+^2  \delta\ell,
    \label{totale}
\end{eqnarray}
which is the thermal energy of photons and pairs,
as will be discussed in the next section. Note that the trapped thermal energy is not small, but comparable with the expanding outflow energy for GRBs, see Fig.~\ref{trapped} caption and Ref.~\cite{Ruffini2003}.
Since the energy density $\rho_\gamma(r)$ of photons and pairs is highly peaked near the horizon, as shown in Fig.~\ref{trapped}, we use the $\delta$-function approximation 
$\int (1-r_+/r)^{-1/2}\rho_\gamma(r)dr\approx \rho_\gamma(r_+)\delta\ell$, and $\delta\ell=(1-r_+/r)^{-1/2}dr$ is a proper distance near the horizon. The macroscopic length $\delta\ell$ should be much larger than the Compton length $\lambda_e$, but smaller than the fireshell size (width) $2G M\approx 2.33 \times 10^6$cm. 

\subsection{Thermodynamic and opaque radiation fluid of photons and pairs}

The photons and pairs' energy (number) density 
$\rho_\gamma$ ($n_\gamma$) is extremely 
large inside the trapped fireshell. The mean free path $\xi_\gamma=(\sigma_\gamma n_\gamma)^{-1}$ and the thermalization time scale $\tau_\gamma=(\sigma_\gamma  n_\gamma)^{-1}c$ are much smaller than the macroscopic fireshell size $\bar R$ and the time scale $\bar R/v_s$. Therefore, the local thermal equilibrium is almost instantly established via the back-and-forth process $\gamma+\gamma\leftrightarrow e^++e^-$ (\ref{bandf}) of the cross section,
\begin{eqnarray}
\sigma_\gamma \approx\sigma_{_T}\frac{3}{8\omega_\gamma}\left[\ln\left(2\omega_\gamma\right)-1\right],\quad \omega_\gamma=\frac{T_\gamma}{m_e}\gg 1,
\label{ggee}
\end{eqnarray}
and time and length scales are
\begin{eqnarray}
\tau_\gamma=\xi_\gamma=(\sigma_\gamma  n_\gamma)^{-1} \approx\left[\sigma_{_T}n_\gamma \frac{3}{8\omega_\gamma}\left[\ln\left(2\omega_\gamma\right)-1\right]\right]^{-1},
\label{ttime}
\end{eqnarray}
where the Thomson cross section $\sigma_{_T}=(8\pi/3)\alpha^2/m^2_e$ and the fine structure constant $\alpha=1/137$. 

\begin{figure}
\centering
\includegraphics[width=3.0in]{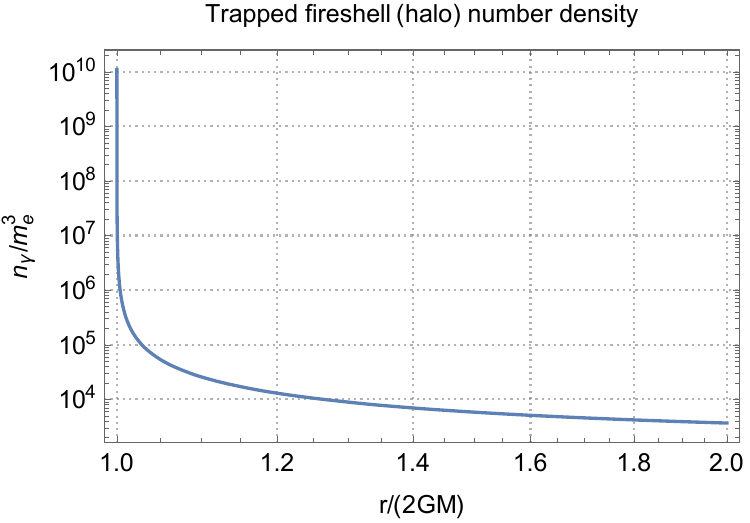}
\includegraphics[width=3.0in]{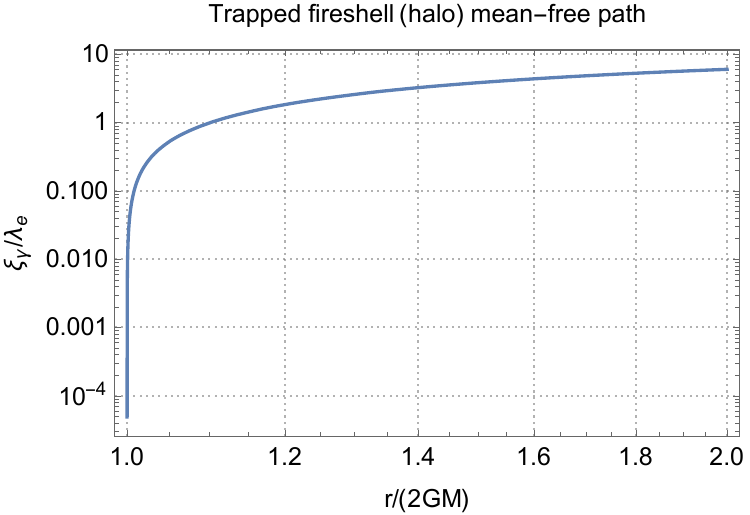}
\caption{The photon and pair number density $n_\gamma$ (left) and mean-free path $\xi_\gamma=(\sigma_\gamma n_\gamma)^{-1}$ (right) are plotted as functions of radius $r$ in the local observer frame. 
The Compton number density $m_e^3= 1.74\times 10^{35}/{\rm cm}^3$ and length $\lambda_e=m_e^{-1}=3.86\times 10^{-11}{\rm cm}$.
}\label{plasma}
\end{figure}

The local temperature $T_\gamma(r)$ represents the mean kinematic energy of photons and electron-positron pairs, which possess zero chemical potential.
As a result, the trapped fireshell is an electrically neutral, opaque and thermalised fluid
of photons and electron-positron pairs, which we also call a radiation fluid for short. 
We describe it by a thermodynamic state of thermal temperature $T_\gamma(r)$, energy and number densities
\begin{eqnarray}
    \rho_\gamma(r)=\sigma_{_{\rm SB}} 
    d_f T_\gamma^4(r),\quad n_\gamma(r)=\frac{30\zeta(3)}{\pi^4}d_f\sigma 
    T_\gamma^3=3.7 d_f \sigma_{_{\rm SB}} T_\gamma^3(r),
    \label{tem}
\end{eqnarray}
in the local frame at the radius $r$. 
The Stefan-Boltzmann constant $\sigma_{_{\rm SB}}=\pi^2/15$ for setting the Boltzmann constant $k_B=1$. The temperature $T_\gamma(r)$ (\ref{tem}), Fig.~\ref{trapped} (right), is the averaged kinetic energy of relativistic particles over species $d_f$. 
In the Standard Model (SM) of
fundamental particle physics, we adopt $d_f\approx 30$ for the number of massless gauge bosons and light 
fermion-antifermion pairs \footnote{Since the number density is so high, see Fig.~\ref{plasma}, neutrinos and anti-neutrinos can be possibly kept inside the fluid in participation of thermalisation.}, which can kinematically 
participate in the thermal equilibration at temperature $T_\gamma> 10$ MeV. The boundary conditions (\ref{bcon}) gives the temperature $T_\gamma^+$ near the horizon,
\begin{eqnarray}
T_\gamma^+\equiv T_\gamma(r_+)=\rho^+_\gamma/(\sigma_{_{\rm SB}} 
    d_f),
\label{bcont}
\end{eqnarray} 
whose value depends on the balance between ultra-relativistic particles and gravitational pressures. 
The horizon temperature $T^+_\gamma$ characterises the maximum temperature 
of a trapped photon-pair fireshell at the place where the gravitational force is maximum. 
For these discussions, the equipartition of particle species' kinetic energies is necessary; an exact thermal equilibrium is not required.

As shown in Figs.~\ref{trapped} and \ref{plasma}, the trapped fireshell's energy, number densities, and temperature are maximal near the horizon and decrease rapidly in radius. It forms a spherical incandescent halo of ultra-relativistic photons and pairs near the horizon. 
The halo width is about $2 GM=1.16\times 10^6$ cm for 
the temperature varying from the maximum $\sim 10^2$ MeV near the horizon to $\sim 1$ MeV near the outer boundary. The halo ends at the outer boundary, where 
electrons and positrons annihilate to two low-energy ($<1$ MeV) photons, and the reverse process is kinematically impossible. The total number of photons and pairs in the fireshell (halo) is enormous 
\begin{eqnarray}
   % N_\gamma&=&\frac{(2GM)^3}{\lambda_e^3}\int_1^2\frac{4\pi r^2dr}{(1-\frac{1}{r})^{1/2}}n_\gamma(r)= 1.14\times 10^8\frac{(2GM)^3}{\lambda_e^3}\gg 1,,\nonumber\\
     N_\gamma&=&\int_{r_+}^{2r_+}n_\gamma(r)\frac{4\pi r^2dr}{(1-\frac{r_+}{r})^{1/2}}\approx n_\gamma^+ 4\pi r_+^2  \delta\ell,\label{totaln}
\end{eqnarray}
corresponding to the total energy $E_\gamma$ (\ref{totale}). Here, the maximal photon-pair number density near the horizon is $n_\gamma^+\equiv n_\gamma(r_+)$,  the macroscopic length $\delta\ell$ is approximately equal to the one in Eq.~(\ref{totale}) for the total energy. Therefore, we effectively describe the trapped fireshells by the temperature $T^+_\gamma$ and length $\delta\ell$.

The proper time and energy of physical processes occurring in the fireshell near the horizon receive significant gravitational dilation and redshift with respect to an observer at infinity, see Fig.~\ref{fallin} and its caption. The fireshell differs from the photon sphere (ring) of radius $3r_+/2$, which describes photons (light) forced to travel in orbits, due to the strong gravitational field near the horizon. 
%photon sphere \cite{Vagnozzi2023}

\subsection{Temperature radial gradient and anisotropic radiation flux}

Photons and pairs in the fireshell are not freely streaming outward, because the mean-free path (\ref{ttime}) is small,
shown in Fig.~\ref{plasma}. On the microscopic scale, a local spacetime around each local point $(r,t)$ inside the fireshell, random particle collisions in all directions, the thermal equilibrium state and temperature $T_\gamma(r)$ are spatially isotropic without 
any preferential direction. 
However, on the macroscopic scale, the temperature  $T_\gamma(r)$ has a large gradient in the radial direction due to the strong gravitational field near the horizon. 
Such a temperature $T_\gamma(r)$ gradient results in a radially outward diffusion of photons and pairs inside 
the fireshell. Following the photon-pair energy density $\rho_\gamma(r)$ and its value (\ref{bcon}) near the horizon, 
we describe the diffusion by the flux $F_\gamma=\rho_\gamma c$ and luminosity $L_\gamma=4\pi r^2\rho_\gamma c$, whose values near the horizon are
\begin{eqnarray}
F_\gamma^+\equiv \rho_\gamma(r_+)c=\rho_nc,~~~L_\gamma^+\equiv L_\gamma(r_+)=\eta L_{\rm pl},
\label{bconf}
\end{eqnarray}
where the ``Planck luminosity'' $L_{\rm pl}\equiv 4\pi r_+^2\rho_{_M}c=(3/2)m^2_{\rm pl}= 3.73\times 10^{59}{\rm ergs/sec}$.

As speculated in Refs.~\cite{Xue_2021, Xue2025}
and will be discussed below, the diffusion flux dissipates the fireshell energy $E_\gamma$ (\ref{totale}) to the blackbody radiation and the emissions of UHE particles. 
Therefore, the fireshell energy density and pressure $P_\gamma=\rho_\gamma/3$ decrease. The fireshell 
is a metastable configuration that gradually shrinks due to gravity. Since the fireshell opacity is large, the energy dissipation rate $\dot E_\gamma/E_\gamma$ is small, and processes are nearly adiabatic. %The energy dissipation time scale $(\dot E_\gamma/E_\gamma)^{-1}$ is large, compared to the fireshell hydrodynamical relaxation time scale of $2r_+/v_s$. 
Thus, we will approximately consider the fireshell as a 
steady-state configuration (source) with a large thermal energy $E_\gamma$ (\ref{totale}) budget and an enormous number of relativistic particles $N_\gamma$ (\ref{totaln}).

In this fireshell example for {\it highly energetic arenas}, we use the TOV equation (\ref{tov00}) and local temperature $T_\gamma(r)$ to describe a macroscopic-sized fireshell of the opaque thermal fluid of dense and energetic photons and pairs, which is trapped around a spherical black hole horizon following a massive stellar core collapse. We ought to mention the following points. 
\begin{enumerate}[(1)]
\item
One can generalise such a thermal fireshell to macroscopic-sized 
firespots in strong gravity and non-spherically symmetric environments of astrophysical accretion and coalescence 
in the presence of angular momentum, magnetic fields and other components. 
\item The gravity strength decrease results in the energy density $\rho_\gamma$ and temperature $T_\gamma$ decreases, the latter is also affected by the number $d_f$ (\ref{tem}) of particle species participating in the thermal equilibrium. 
When the temperature (mean kinetic energy) drops below the hadronic binding energy, hadronisation occurs, and 
the photon-lepton(pair) and gluon-quark(pair) thermal
configuration becomes the photon-lepton(pair) one with electrons, protons and neutrons. 
\end{enumerate}

As will be discussed below, the dynamics producing UHE particles are on microscopic scales $\sim 10^{-2}$ cm, independently of firespots' macroscopic shapes and 
global symmetries. The essentially 
relevant quantities for the dynamics are maximum temperature $T_\gamma^+ > 10$ MeV, the macroscopic size $\delta\ell$ (\ref{totale}) and baryon matter contamination $B$ of the firespots.

\subsection{Baryon matter contamination, bulk electrons and protons}

The trapped fireshell of photons and pairs 
around the black hole horizon has an ultra-high temperature 
$T_\gamma$, see Fig.~\ref{plasma}, which, however, drops quickly below $100$ MeV as the radius increases from the horizon. The quark-gluon plasma undergoes a phase transition of hadronisation and becomes a baryon matter, some of which contaminates the radiation fluid of photons and pairs inside the fireshell. 
We are interested in the temperature range $T_\gamma\sim (10\sim 50)$ MeV, and   
the baryon matter inside the fireshell is fully ionised, comprising electrons ($e$) and protons ($p$), which are electrically neutral in total.

These electrons ($e$) and protons ($p$) have to be distinguished from electrons $e^-$ and positrons $e^+$ associated  
with two photons $2\gamma$ (\ref{bandf}). They
have non-zero chemical potentials and do not participate in the back-and-forth process $\gamma+\gamma\leftrightarrow e^++e^-$ (\ref{bandf}) for the thermal equilibrium of photons and pairs. Therefore, their mean kinetic energies do not share with the mean kinetic energies (temperature $T_\gamma$) of photons and pairs.
These electrons ($e$) and protons ($p$) effects on the thermal equilibrium, which were studied by using the Boltzmann equation in Ref.~\cite{Ruffini2000}, are small if their average energy and number densities are much smaller than those of photons and pairs.

Thus, we model such an ionised baryon matter contamination as dilute electron ($e$) gas and proton ($p$) gas immersed inside the thermal and neutral radiation fluid of photons and pairs at the temperature $T_\gamma$. 
In this article, we focus on these electrons ($e$) and protons ($p$), their electromagnetic interaction with a dense radiation fluid of photons and pairs. We ignore neutrons and weak interactions associated with the inverse $\beta$ decay and anti-neutrino production.
 
Colliding with photons and pairs via multi-particle collisions and field interactions, these electrons ($e$) and protons ($p$) are in microscopic random motions in all directions. They form a relativistic ($e$) gas and a non-relativistic ($p$) gas, and do not form an electron-proton oscillating plasma.
We describe them by {\it bulk} electrons and protons inside the fireshell, whose numbers are, 
\begin{eqnarray}
\bar N_e &=& \bar N_p = B N_\gamma.
\label{bulkb}
\end{eqnarray}
Here, the baryon-loading $B\ll 1$ quantifies the total number of electrons (protons) inside an opaque fireshell of photons and pairs of the number $N_\gamma$. 
The $B$ should relate to the baryon loading of fireballs, whose outward expansion accounts for GRBs. 

We adopt the notations $\tilde\beta_e$ and $\tilde\gamma_e$ to represent the velocity and Lorentz factor of bulk electrons \footnote{Bulk electrons (protons) are relativistic (non-relativistic), which indicates individual bulk electron (proton) of Lorentz factor $\tilde\gamma_e>1$ ($\tilde\gamma_p\gtrsim 1$) without being averaged over their distribution function in energy-momentum phase space. It is analogous to an opaque radiation fluid of relativistic photons and pairs, whose macroscopic diffusion (drifting) velocities are small, corresponding to an averaged Lorentz factor close to unity.}, 
which are in random motions at microscopic scales, therefore,
\begin{eqnarray}
\langle \tilde \beta_e\rangle=0,~~~ \langle \tilde \gamma_e\rangle =1,
\label{bulk}
\end{eqnarray}
where the averages $\langle \tilde \beta_e\rangle$ and $\langle \tilde \gamma_e\rangle$ are over the energy-momentum distribution of bulk electrons. The average values (\ref{bulk}) mean that bulk electrons have no macroscopic motions in any preferential direction. However, due to an external radiation force or other dynamics at macroscopic scales, the average values can be non-zero: $\langle \tilde \beta_e\rangle\not=0$ and $\langle \tilde \gamma_e\rangle>1$. It implies that some of the bulk electrons tend to macroscopic motions in a preferential direction and with a non-vanishing velocity  
\begin{eqnarray}
\beta_e=\langle \tilde \beta_e\rangle\not=0,~~~ \gamma_e= \langle \tilde \gamma_e\rangle >1.
\label{bulkv}
\end{eqnarray}
We call them (\ref{bulkv}) {\it drifting} electrons with a non-vanishing velocity $\beta_e\not=0$ and Lorentz factor $\gamma_e>1$ at macroscopic scales. 
The same discussions apply to bulk and drifting protons. We now turn to discuss the dynamics of radiation force accelerating electrons from the bulk electrons (\ref{bulk}) to the drifting electrons (\ref{bulkv}).

\section{Compton-rocket effect on accelerating electrons}\label{Compton}

\subsection{Radiation force acting on bulk electrons}

Reference \cite{ODell1981} studies the Compton-rocket effect: via Thomson scattering, the radially outward photon flux (intensity) $F_\gamma(r)=\rho_\gamma(r)c$ acts as a radiation force on an optically thin gas of hot electrons or cold protons, and accelerates them 
to high energies. The photon-scattering excess force on an isotropic gas results 
from the globally anisotropic loss of internal radiation energy.  
In the one-dimensional model with respect to 
a local observer's proper time $\tau$ in a locally flat Lorentz frame at the radius $r$, the rates of energy and momentum transfer from the radiation flux $F_\gamma(r)$ to the electron gas are (per electron),
\begin{eqnarray}
\frac {d\epsilon_e}{d\tau} &=&-\sigma_{_T}F_\gamma(r)[(4/3) (\beta_e \gamma_e)^2],\label{acce0}\\
\frac {d p_e}{d\tau} &=& \sigma_{_T} F_\gamma(r)[(2/3)(\beta_e \gamma_e)^2+1],
\label{accp0}
\end{eqnarray} 
and some of the bulk electrons\footnote{The phrase ``some of the bulk electrons'' will be quantified by the fraction $N_e/\bar N_e$ (\ref{driftf}), 
which equivalently corresponds to the probability of a bulk electron accelerated to a drifting electron in the macroscopic motion along the radial direction.} $\bar N_e$ are accelerated to drifting electrons $N_e$ with macroscopic velocity $\beta_e=v_e/c$, Lorentz factor $\gamma_e=(1-\beta^2_e)^{1/2}$, and 
momentum $p_e=m_e\beta_e\gamma_e$, $(\beta_e\gamma_e)^2=\gamma_e^2-1$, corresponding to (\ref{bulkv}).
The initial conditions  at $\tau=0$ for these dynamical equations (\ref{acce0}) and (\ref{accp0}) are $\beta_e=\langle \tilde \beta_e\rangle=0$ and $\gamma_e=\langle \tilde \gamma_e\rangle =1$ (\ref{bulk}), when electrons are bulk electrons. Note that the radiation flux $F_\gamma(r)$ due to multi-photon scattering off an electron should be considered as a collective force at macroscopic scales, rather than a single photon scattering off an electron at microscopic scales. 

To simplify the notations consistently with the one-dimensional model, we do not specify the averages $\langle\cdot\cdot\cdot\rangle$ 
over the electron distribution
in the energy-momentum phase space.
Equation (\ref{accp0}) gives the Eddington luminosity argument for the incident radiation pressure on 
non-relativistic electrons ($\beta_e\approx 0$) 
in the Thomson scattering limit \cite{ODell1981}. 
For the proton case, these formulae and discussions remain the same with the substitutions of the subscript 
$e\rightarrow p$ and approximate cross section 
$\sigma_{_T}\rightarrow \sigma^p_{_T}\approx (m_e/m_p)^2\sigma_{_T}\ll \sigma_{_T}$ of multi-photons and proton collisions. As such, the Compton-rocket effect on protons is much smaller than on electrons.

Radiation fields non-linearly interact with  relativistic electrons via the Compton rocket effect \cite{ODell1981, Sikora1981, Phinney1982} 
by the anisotropy of multi-photons and electron scattering. This dynamics has been studied in an optically thin plasma of electrons and protons by particle-in-cell numerical simulations \cite{Frederiksen_2008, DelGaudio2020, Martinez2021, Faure2024}.
It is worthwhile to emphasise that in the Compton-rocket effect (\ref{acce0},\ref{accp0}), the photon flux $F_\gamma(r)$ is essentially attributed to the nonlinear QED processes
of multi-photon scatterings and interactions (see, e.g., \cite{2014PhRvD..90a3009W, Zhang:2018jje, Zhang2019}). 
These non-linear interactions are particularly important when the radiation field is energetic and the charged particle density is large. It is what we will study in the following sections.

\subsection{Drifting electrons in an opaque radiation fluid at high temperatures}

In the one-dimensional and spherical symmetric fireshell, 
the bulk electrons averaged momentum $\langle\tilde p_e\rangle= m_e\langle\tilde\beta_e\tilde \gamma_e\rangle$ vanishes 
for the random momentum distribution $\propto e^{-\tilde p_e^2/T_\gamma}$ with an equal probability of inward and outward moving. Thus, bulk electrons $\bar N_e$ do not have a net drift motion in the direction $\hat r>0$ or opposite one $\hat r<0$, and their averaged velocity vanishes $\langle\tilde\beta_e\rangle=0$ 
and $\langle\tilde\gamma_e\rangle=1$ (\ref{bulk}), 
and the same discussions for bulk protons $\bar N_p$.

However, the situation changes when the strong radiation force (\ref{accp0}) is present, acting on bulk electrons, since the density of photons and pairs has a large variation gradient 
in the radial direction, as shown in Figs.~\ref{trapped} and \ref{plasma}. 
We attempt to investigate the Compton-rocket effect on bulk electrons 
in a highly opaque fluid of photons 
and pairs at high temperatures $T_\gamma\gg m_e$, 
densities $\rho_\gamma \gg m_e^4$ and $n_\gamma \gg m_e^3$. 
Namely, how the radially outward flux (intensity) $F_\gamma(r)=\rho_\gamma(r)c\gg m_e^4$ acts as a radiation force on bulk electrons and protons (\ref{bulk}), 
accelerating {\it some of them} outward along the radial direction to ultra high energies, drifting out of the fireshell.

This system is more complex than 
an optically thin electron-proton plasma irradiated at energies $\epsilon_\gamma\sim m_e$ photons, numerically studied, 
see e.g.~Refs.~\cite{Faure2024,2013CoPhC.184.2503X}. Nevertheless, to gain insight into the dynamics of the Compton-rocket effect on bulk electrons and protons immersed in the dense, energetic radiation fluid of photons and pairs, we use a simplified one-dimensional model along the radial direction only. The photon and pair flux $F_\gamma=\rho_\gamma(r)c$ acts on bulk electrons and protons in a tiny tube with the cross section $\sigma_\gamma$.

\subsection{Drifting electrons versus bulk electrons}

Although the mean-field path of bulk electrons is small, as shown in Fig.~\ref{plasma}, the outward radiation force $F_\gamma=\rho_\gamma(r)c$ is strong, so some bulk electrons have chances within their mean-free path to get accelerated 
and drift radially outwards. 
Therefore, some of bulk electrons drift in the direction $\hat r>0$ and their averaged momentum $p_e=\langle\tilde p_e \rangle_a$ 
and energy $\epsilon_e=(p_e^2+m_e^2)^{1/2}$ do not vanish. 
They are {\it drifting} electrons with the macroscopic velocity $\beta_e$ and Lorentz factor $\gamma_e$ (\ref{bulkv}), and we define the number of 
these outward drifting electrons as 
\begin{eqnarray}
N_e (\gamma_e)< \bar N_e,~~~ \beta_e = \langle \tilde \beta_e\rangle_a >0, ~~~  \gamma_e= \langle \tilde \gamma_e\rangle_a >1,
\label{drift}
\end{eqnarray}
where the average $\langle\cdot\cdot\cdot\rangle_a$ is over the energy-momentum distribution of bulk and drifting electrons. The drifting electrons $N_e (\gamma_e)$ should be a small fraction of the bulk electrons $\bar N_e$, because most of the radiation force accelerated electrons 
do not become drifting electrons, but run back to bulk electrons due to collisions with photons and pairs. 
We define the fraction (radio) of drifting and bulk electrons as
\begin{eqnarray}
\frac{N_e}{\bar N_e}\equiv \frac{N_e(\gamma_e)}{\bar N_e},   
\label{driftf}
\end{eqnarray}
which actually represents the probability that electrons are accelerated out of bulk electrons to become drifting electrons by the Compton-rocket effect in an opaque photon-pair fluid. This fraction (\ref{driftf}) is a crucial quantity we adopt to describe the significance of the Compton-rocket effect on accelerating electrons to a Lorentz factor $\gamma_e$ in an opaque photon-pair fluid.
The same discussions apply to drifting $N_p$ 
and bulk $\bar N_p$ protons in the 
opaque fluid of photons and pairs.

We further examine the time scale $\tau_a^e$ of the electron acceleration process (Compton-rocket effect) 
and compare it with the thermalisation time scale $\tau_\gamma=(\sigma_\gamma n_\gamma)^{-1}c$ (\ref{ttime}) of photon and pair fluid. Using Eq.~(\ref{accp0}) we define 
\begin{eqnarray}
\tau_a^e\equiv \frac{p_e}{dp_e/d\tau}=\frac{m_e\beta_e\gamma_e}{\sigma_{_T}F_\gamma [\frac{2}{3}(\beta_e\gamma_e)^2+1]},
\label{atime}
\end{eqnarray}
and obtain the ratio
\begin{eqnarray}
\frac{\tau_a^e}{\tau_\gamma}=\frac{3.7\times 9}{16\omega^2_\gamma}\frac{\ln\left(2\omega_\gamma\right)-1}{(\beta_e\gamma_e)\left[1+\frac{3}{2(\beta_e\gamma_e)^2}\right]}.
\label{freer}
\end{eqnarray}
Here, the gravitational time dilation factors $d\tau=(1-r_+/r)^{1/2}dt$ are simplified. Recall that the thermalisation time scale $\tau_\gamma$ (\ref{ttime}) represents also the time scale of the radiation flux $F_\gamma$ of multi-photons acting on a changed particle.

The time and length scales $\tau_a^e$ of the Compton-rocket effect are smaller than the 
thermalisation timescale $\tau_\gamma$ of photons and pairs 
fluid, i.e., $\tau_a^e<\tau_\gamma$, for $\omega_\gamma =T_\gamma/m_e > 1$ and $\gamma_e> 1$. The drifting electron density $n_e(\gamma_e)$ is much 
smaller than the density $n_\gamma$ of photons and pairs.
These properties imply that drifting electrons accelerated by the Compton-rocket effect are small and rapid perturbations upon the thermal bath of photons and pairs at high energies 
and densities. If the timescale inequality was 
$\tau_a^e>\tau_\gamma$ instead of $\tau_a^e<\tau_\gamma$, 
the perturbations of drifting electrons would have been washed out by photon-pair thermal fluctuations. Note that the drifting electron perturbation's energy and momentum are $\delta \epsilon_e >m_e$ and $\delta p_e>m_e$ with Lorentz factor $\gamma_e\gtrsim 1$. 

Moreover, due to the large opacity and frequent collisions with photons and pairs, drifting electrons ($\gamma_e\gtrsim 1$) quickly dissipate their energy and run back to bulk electrons ($\gamma_e=1$), which are in an energy equipartition with photons and pairs. It implies 
that the perturbations of drifting electrons ($\gamma_e>1$) should be rapidly washed out. 
Therefore, it looks unlikely that in the opaque fluid of photons and pairs, such small perturbations could develop, 
namely, some accelerated electrons by the Compton-rocket effect could drift to 
a large Lorentz factor $\gamma_e\gg 1$, unless some instabilities of avalanche phenomena occur.
 
\section{Electrons runaway in an opaque radiation fluid}\label{run}

\subsection{Instability and electron runaway 
due to Klein–Nishina corrections}\label{knsec}

As discussed, in the opaque fluid of photons and pairs, the 
Compton-rocket effect on accelerating electrons is an
approximately ``local'' and ``instantaneous'' perturbation that triggers electrons to increase energy and drift in the outward radial direction. On their way, accelerated electrons collide with photons and pairs with the probability described by the 
Klein–Nishina (KN) cross section (\ref{knh}), which, however, decreases as the accelerated electrons' energy increases, and the straightforward scattering with a narrower deflection angle $\theta_e\sim 0$. 
This property of KN corrections implies the instability of an unstable avalanche runaway process of drifting electrons. Namely, accelerated electrons gain more energy and have fewer collisions. Thus, they have greater chances to 
drift a longer distance and gain even more energy from the Compton-rocket effect. As a result, there is a probability that a small fraction of drifting electrons run away and achieve ultra-high energies, 
penetrating through the opaque fluid of photons and pairs. 

In the opaque fluid of photons and pairs at the temperature $T_\gamma >m_e$, drifting electrons move outward in the radial direction
$\theta_e\sim 0$, 
with the velocity $\beta_e\sim 1$ and Lorentz factor $\gamma_e> 1$. 
The KN cross-section is approximately given by 
\begin{eqnarray}
\sigma^e_{_{\rm KN}} \approx\sigma_{_T}\frac{3}{8\omega_e}\left[\ln\left(2\omega_e\right)+\frac{1}{2}\right],\quad 
\omega_e\approx \frac{2T_\gamma}{m_e}\gamma_e,
\label{knh}
\end{eqnarray}
where the factor $2$ in $\omega_e$ (\ref{knh}) is due to the blue-shift effect \cite{1974ApJ...188..121B}. 
The cross-section decreases as the drifting electrons' Lorentz factor $\gamma_e$ increases and vanishes when $\gamma_e \gg 1$. Therefore, high $\gamma_e$ drifting electrons have a small probability of being scattered back to bulk electrons with $\gamma_e=1$. Instead, they are further accelerated to higher $\gamma_e$ by the Compton-rocket effect, and behave as runaway drifting electrons. 

Such a runaway process 
has a time scale $\tau^e_{_{\rm KN}}=(\sigma^e_{_{\rm KN}}n_\gamma c)^{-1}$ of the instability, 
which is much larger than the thermalization time scale $\tau_\gamma$ (\ref{ttime}) of photons and pairs,
\begin{eqnarray}
\frac{\tau^e_{_{\rm KN}}}{\tau_\gamma}\approx2\gamma_e\frac{\ln\left(2T_\gamma/m_e\right)}{\ln\left(4\gamma_eT_\gamma/m_e\right)}\gg 1,~~~ \gamma_e\gg 1,
\label{kntime}
\end{eqnarray}
here the gravitational time dilation factors $d\tau=(1-r_+/r)^{1/2}dt$ are simplified. The time scales (\ref{atime}), (\ref{freer}) and (\ref{kntime}) give 
$\tau^e_a<\tau_\gamma \ll \tau^e_{_{\rm KN}}$, 
implying that rather than being smeared out, the instability could develop a runaway solution. It makes drifting electrons start from microscopic perturbations (\ref{freer}) and run to macroscopic 
motions (\ref{kntime}). 

\subsection{Probability of drifting electrons runaway}
Even if such an instability occurs, in view of drifting electrons colliding with photons and pairs, what is the probability that runaway drifting electrons survive when they travel over a long macroscopic distance and time scale, reaching a large Lorentz factor $\gamma_e$? 
We describe this probability by the fraction $N_e/\bar N_e$ (\ref{driftf}) of drifting and bulk electron numbers, and study it for the following three cases:
\begin{enumerate}[(i)]
\item $N_e/\bar N_e=1$ (maximal probability) corresponds 
to all of the bulk electrons being
accelerated and becoming drifting electrons at Lorentz factor $\gamma_e\gg 1$. This collisionless case can not be realistic, since the drifting 
electrons dissipate their kinetic energies, and return to bulk electrons via collisions with photons and pairs in the fireshell of a large opacity. Nevertheless, we study this case to gain an insight into the nature of the Compton-rocket effect on accelerating electrons inside an opaque fluid of energetic and dense photons and pairs.
\item $N_e/\bar N_e\approx 0$ (near zero probability) corresponds 
to almost none of the bulk electrons being
accelerated and becoming drifting electrons at Lorentz factor $\gamma_e\gg 1$. In other words, almost all drifting electrons return to bulk electrons due to collisions with photons and pairs. It indicates that the instability does not develop into an avalanche runaway
process, and a negligible amount of bulk electrons are
accelerated and become drifting electrons at Lorentz factor $\gamma_e\gg 1$. It is the collisional case of Thomson scattering 
with an energy-independent cross section at low energies.
\item $N_e/\bar N_e\not=0$ is a non-trivial value   
(finite probabilities), corresponding 
to a significant amount of the bulk electrons being
accelerated and becoming drifting electrons at Lorentz factor $\gamma_e\gg 1$. It indicates that the instability has developed into an avalanche runaway
process, and a significant amount of bulk electrons are
accelerated and become drifting electrons at Lorentz factor $\gamma_e\gg 1$.
It is the collisional case of Klein–Nishina scattering 
with an energy-dependent cross section at high energies.
\end{enumerate} 
To quantitatively study this fraction $N_e/\bar N_e$ in these cases, we need two master equations for drifting electrons: (1) the acceleration equation 
(\ref{accp1}) attributed to the Compton-rocket effect, (2) the decay rate equation (\ref{prob10}) 
or (\ref{prob1}) due to drifting electrons' energy-momentum dissipation in collisions with photons and pairs inside the fireshell.

We rewrite the Compton-rocket effect on the electron acceleration (\ref{accp0}) in a small distance 
$dr=\beta_e d\tau=\beta_e(1-r_+/r)^{1/2}dt$ as 
\begin{eqnarray}
m^{\rm eff}_e\gamma_e\frac{d\gamma_e}{dr} &=& \frac{1}{3}\sigma_{_T} (1-r_+/r)^{-1/2}F_\gamma(r)(2\gamma_e^2+1).
\label{accp1}
\end{eqnarray}
In the one-dimensional model, this equation represents  
drifting electrons $N_e(\gamma_e)$ of $\gamma_e>1$ accelerated out of bulk electrons $\bar N_e$ of $\gamma_e=\langle\tilde \gamma_e\rangle =1$. The boundary condition is $\gamma_e=1$ at $r=r_+$ 
from the initial condition for (\ref{accp0}).
The effective
mass $m^{\rm eff}_e=m_p+m_e\approx m_p$ will be explained in due course.

\subsection{Electrons accelerated by an opaque radiation fluid at high temperature}

To have an idea about the spacetime length scales of electron acceleration in the fireshell of energetic and dense photons and pairs, we {\it first} consider the collisionless case where all bulk electrons 
$\bar N_e$ (\ref{bulkb}) are accelerated to drifting electrons $N_e(\gamma_e)$ 
in the radial direction without any collision with photons 
and pairs. Then integrating 
the Compton-rocket effect (\ref{accp1}) yields
\begin{eqnarray}
2\gamma_e^2 +1= 3 \exp \Big[\frac{4\sigma_{_{T}}}{3m^{\rm eff}_e}\int_{r_+}^r \frac{\rho_\gamma(r')}{(1-r_+/r')^{1/2}}dr'\Big],
\label{free}
\end{eqnarray}
where bulk electrons $\gamma_e=1$ near the horizon $r_+$ are accelerated to drifting electrons $\gamma_e>1$ at the radius $r$.
The photon and pair flux 
$F_\gamma(r)=\rho_\gamma c$ is highly peaked and has a great gradient near the horizon $r_+$,
as shown in Fig.~\ref{trapped}. Therefore, the photon and pair flux 
$F^+_\gamma=\rho_\gamma(r_+)c$ (\ref{bconf}) and temperature $T^+_\gamma=T_\gamma(r_+)$ (\ref{bcont}) near the horizon 
predominantly contribute to the integration (\ref{free}), i.e., to the acceleration of drifting electrons out of bulk electrons. Similarly to Eq.~(\ref{totale}) for the total energy, 
we use the $\delta$-function approximation $\int (1-r_+/r)^{-1/2}\rho_\gamma(r)dr\approx \rho_\gamma(r_+)\delta\ell=\sigma_{_{\rm SB}}d_f T^4_\gamma(r_+)\delta\ell$, where $\delta\ell=(1-r_+/r)^{-1/2}dr$ is a small proper distance near the horizon. Solution (\ref{free}) approximately becomes
\begin{eqnarray}
2\gamma_e^2 +1\approx  3 \exp \Big[\frac{4\sigma_{_{T}}\sigma_{_{\rm SB}}d_f}{3m^{\rm eff}_em_e}T^4_\gamma(r_+)\delta\ell\Big],
\label{freet}
\end{eqnarray}
namely, the radiation force $F_\gamma(r_+)=\rho_\gamma(r_+)c$ near the horizon, where the energy density $\rho_\gamma(r_+)$ and temperature $T^+_\gamma$ are maximal, accelerates on an electron over a distance $\delta\ell$ to reach its Lorentz factor $\gamma_e$. 

In the viewpoint of the energy density $\rho_\gamma\sim T_\gamma^4 \sim E^2_{\rm eff}/(8\pi)$,
the photons and pairs radiation force (pressure) $F_\gamma=c\rho_\gamma$ for the temperature $T_\gamma^+\equiv T_\gamma(r_+)\sim {\mathcal O}(10^{1-2})$MeV 
is equivalent to an effective 
electric force  $eE_{\rm eff}$, which is about
${\mathcal O}(10^{1-2})$ times larger than the critical electric field force $eE_c=m_e^2$ for electron-positron pair productions. 
In a classical scenario of a test charged particle accelerated by an electric field, we estimated that 
such an effective electric force $eE_{\rm eff}$ accelerates an electron to PeV $\sim 10^9 m_e (\gamma_e\sim 10^9)$ energies
in a small distance $d\sim {\mathcal O}(10^9)\lambda_e\sim {\mathcal O}(10^{-2})$ cm
and time $\tau=d/c\sim {\mathcal O}(10^{-12})$ sec \cite{Xue_2021}. 
Therefore, the macroscopic length
$\delta\ell$ determined from the total thermal energy $E_\gamma$  (\ref{totaln}) should be large enough for an electron reaching $\gamma_e\sim 10^9$ in Eq.~(\ref{freet}). 
It indicates that the macroscopic sizes and geometries of dense and energetic %incandescent 
photon-pair fireshells or spots are irrelevant for the Compton-rocket effect, because the local inhomogeneous density of photons and pairs accelerates electrons on microscopic time and length scales $d$, which are much smaller than the macroscopic time and size variation of fireshells or firespots.

Physically, the collisionless solution (\ref{freet}) means that (i) all bulk  
electrons $\bar N_e$ of $\gamma_e=1$ accelerated to drifting electrons $N_e(\gamma_e)$ of an enormous Lorentz factor $\gamma_e\gg 1$; (ii) the majority of drifting electrons get accelerated and reach their ultra-high energies in the neighbourhood near the horizon, or where photon-pair energy density and temperature are maximal.
This collisionless solution (\ref{freet}) of $N_e(\gamma_e)=\bar N_e$ is certainly not realistic, 
since the radiation fluid of photons and pairs is extremely opaque, and drifting electrons 
colliding with photons and pairs is unavailable. However, the study of the collisionless case (\ref{freet}) let us learn (i) 
the large-temperature $T^+_\gamma$ 
dominates the radiation-force contribution to electron acceleration; (ii) the spacetime length scales $d$ of electron acceleration in an opaque radiation fluid of photons and pairs. This helps make appropriate approximations in studying collisional cases, in which the drifting electron cannot reach $\gamma_e\gg 1$ (\ref{freet}) with the maximal probability (fraction) $N_e/\bar N_e=1$.

For the collisional case, to show the runaway dynamics due to the Klein-Nishina correction in high energies, as previously speculated in section (\ref{knsec}), we attempt to calculate the fraction
$N_e/\bar N_e$ numerically in two separate cases: (1) the stable case, where drifting electrons undergo Thomson scattering in low energies; (2) the unstable case, where drifting electrons undergo Klein-Nishina scattering in high energies. We do not adopt the full scattering cross-section that varies with energy. The reasons are that (i) there is no analytical solution $N_e/\bar N_e$ for the entire energy range; (ii) we have numerical difficulties to tackle the runway dynamics when it occurs, 
the fraction $N_e/\bar N_e$ varies by many orders of magnitude, as will be shown in Figs.~\ref{therm500} and \ref{notherm}; (iii) it is helpful to distinctly compare and contrast Thomson and  Klein-Nishina scattering cases 
to understand how the runaway dynamics physically develops.

%\subsection{Superthermal spectrum of UHE electrons} 

\subsection{Stable case: Thomson scattering at low energies}

We consider the collisional case of the constant Thomason scattering cross section $\sigma_{_T}$ at low energies (temperatures). 
We calculate the number of drifting electrons accelerated in one proper mean-free path $\xi_e^{\rm T}= (n_\gamma \sigma_{_T})^{-1}$ and time $d\tau=\xi_e^{\rm T}/\beta_e$. 
The acceleration equation (\ref{accp0}) or (\ref{accp1}) for the 
Lorentz factor variation $\Delta \gamma^{\rm T}_e$ in one mean-free path becomes 
\begin{eqnarray}
m^{\rm eff}_e\gamma_e\Delta \gamma^{\rm T}_e &=& \frac{1}{3}\sigma_{_T} F_\gamma(r)(2\gamma_e^2+1) \xi_e^{\rm T}.
\label{scat0}
\end{eqnarray}
Using the relation $F_\gamma/n_\gamma= T_\gamma/3.7$, the variation $\Delta\gamma^{\rm T}_e$ in one mean-free path is given by
\begin{eqnarray}
\Delta \gamma^{\rm T}_e &=& \frac{1}{3.7}\left(\frac{T_\gamma}{m^{\rm eff}_e}\right)\frac{2\gamma_e^2+1}{3\gamma_e},
\label{dgamma0}
\end{eqnarray}
which is proportional to $(T_\gamma/m^{\rm eff}_e)\gamma_e$ for $\gamma_e\gg 1$.

Due to the Thomson scatterings with photons and pairs, the drifting electron number $N_e(\gamma_e)$ decreases following the one-dimensional rate equation, 
\begin{eqnarray}
\frac{dN_e}{dr} &=& -\frac{1}{\xi_e^{\rm T}}N_e=-n_\gamma\sigma_{_T}N_e,
\label{prob10}
\end{eqnarray}
similarly to the equation for unstable particles of the number $N_e$ decaying while travelling along the radial trajectory.
The rate equation (\ref{prob10}) recasts as
\begin{eqnarray}
\frac{d N_e(\gamma_e)}{d\gamma_e} &=& -\frac{1}{\Delta\gamma^{\rm T}_e}N_e(\gamma_e),
\label{prob20}
\end{eqnarray}
by using Eqs.~(\ref{accp1}) and (\ref{scat0}). 

\begin{figure}
\centering
\includegraphics[width=3.0in]{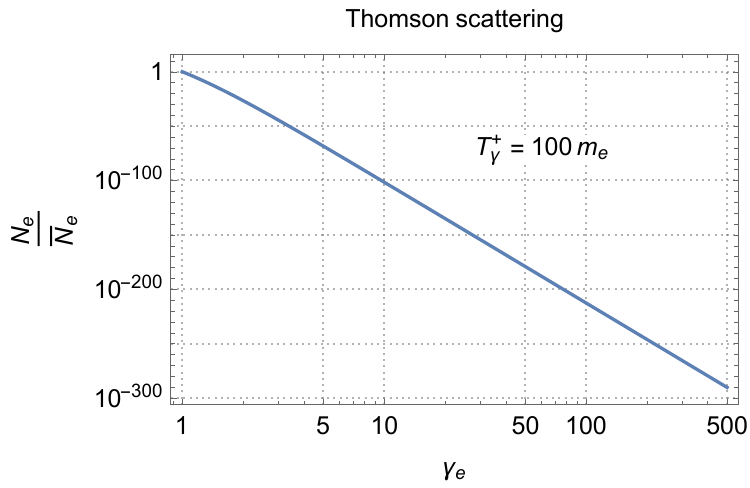}
\includegraphics[width=3.0in]{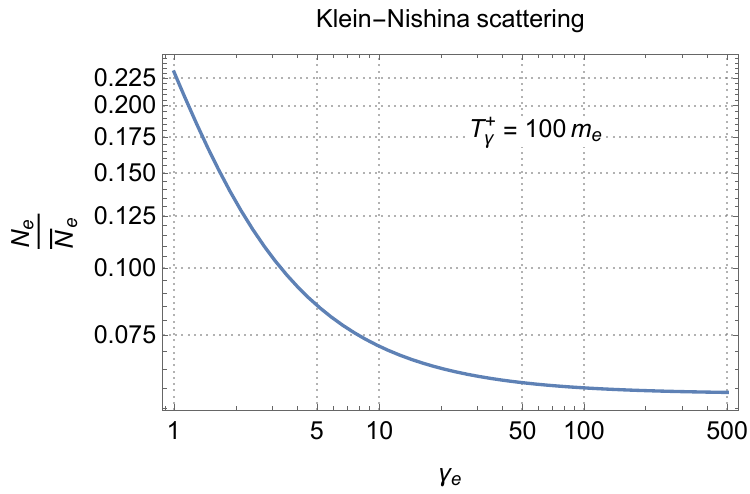}
\caption{We calculate the fraction (probability) $N_e/\bar N_e$ of accelerated electrons drifting out of the bulk electrons for the Thomson (\ref{prob2i0}) and the Klein-Nishina (\ref{prob2i}) scattering cases. It shows that the probability of accelerated electrons reaching high energies is non-trivial, implying the runaway behaviour, in the Klein-Nishina scattering case, but is completely negligible in the Thomson scattering case.   
These figures also represent the proton case with the substitution of the subscript $e \rightarrow p$. }\label{therm500}
\end{figure}

Assuming that the majority of drifting electrons get accelerated in the neighbourhood near the horizon, as discussed in Eq.~(\ref{freet}), we approximate the temperature $T_\gamma\approx T^+_\gamma$ by the horizon temperature $T^+_\gamma$. 
Integrating the rate equation (\ref{prob20}) shows the fraction (probability) of drifting electrons $N_e(\gamma_e)$ out of bulk electrons $\bar N_e$ ($\gamma_e=1$) is given by
\begin{eqnarray}
\frac{N_e(\gamma_e)}{\bar N_e} &=& \exp\left( -\int_1^{\gamma_e} \frac{1}{\Delta\gamma^{\rm T}_e} {d\gamma_e}\right)\approx \left(\frac{3}{2\gamma_e^2+1}\right)^{2.78(m^{\rm eff}_e/T^+_\gamma)}.
\label{prob2i0}
\end{eqnarray}
As shown on the left of Fig.~\ref{therm500}, 
the fraction $N_e(\gamma_e)/\bar N_e$ decreases rapidly as the accelerated electron Lorentz factor $\gamma_e$ increases on the way they travel outward, 
dissipating their energy and momentum into the fluid via collisions with photons and pairs due to the large opacity.

Thus, the most accelerated drifting electrons $N_e(\gamma_e)$ at high energies run back to the bulk electrons $\bar N_e$ at low energies, 
which are in the energy equipartition 
with photons and pairs. In other words, the probability of drifting electrons reaching a large Lorentz factor $\gamma_e\gg 1$ goes to zero at low temperature 
$T^+_\gamma \ll m^{\rm eff}_e$. Very few electrons are in the high-energy superthermal spectrum tail; thus, they are negligible and irrelevant for observations of UHE particles. It is the case for the Sun, whose core temperature of photons is much smaller than the electron mass, i.e., $T_\odot\approx 1.5\times 10^4 {\rm eV} \ll m_e$ and the fraction (\ref{prob2i0}) is
exponentially suppressed.   

\subsection{Unstable case: Klein-Nishina scattering at high energies}
\label{knsca}

We investigate the collisional case of Klein-Nishina scattering at high energies (temperatures), and show a non-trivial probability of drifting electrons running away, reaching $\gamma_e\gg 1$, due to the energy-dependent nature 
of the KN cross section $\sigma^e_{_{\rm KN}}(\gamma_e)$ (\ref{knh}), which decreases with increasing energies. 

Analogously to $\Delta\gamma^{\rm T}_e$ (\ref{scat0}), define $\Delta\gamma^{\rm KN}_e$ to be the Lorentz factor variation of drifting electrons accelerated in one proper mean-free path $\xi_e^{\rm KN}=(n_\gamma \sigma^e_{_{\rm KN}})^{-1}$.
The acceleration equation (\ref{accp0}) or (\ref{accp1}) for the 
Lorentz factor variation $\Delta \gamma^{\rm KN}_e$ in one mean-free path $\xi_e^{\rm KN}$ becomes 
\begin{eqnarray}
m^{\rm eff}_e\gamma_e\Delta \gamma^{\rm KN}_e &=& \frac{1}{3}\sigma_{_T} F_\gamma(r)(2\gamma_e^2+1) \xi_e^{\rm KN}.
\label{scat}
\end{eqnarray}
Using the relation $F_\gamma/n_\gamma= T_\gamma/3.7$ and the KN cross section (\ref{knh}) that depends on drifting electrons' energy, i.e., Lorentz factor $\gamma_e$, we obtain the variation $\Delta\gamma^{\rm KN}_e$ in one mean-free path given by
\begin{eqnarray}
\Delta \gamma^{\rm KN}_e(T_\gamma,\gamma_e) &=& \frac{1}{3.7}\frac{16}{9}\left(\frac{T^2_\gamma}{m_e m^{\rm eff}_e}\right)\frac{2\gamma_e^2+1}{\ln(4\gamma_e\frac{T_\gamma}{m_e})+\frac{1}{2}},
\label{dgamma}
\end{eqnarray}
in the contrast with $\Delta \gamma^{\rm T}_e$ (\ref{dgamma0}) for the Thomason scattering case.

Analogously to Eqs.~(\ref{prob10}) and (\ref{prob20}) in the Thomson scattering case, due to the Klein-Nishina scatterings with photons and pairs, the drifting electron number $N_e(\gamma_e)$ decreases following the one-dimensional rate equation 
\begin{eqnarray}
\frac{dN_e}{dr} &=& -\frac{1}{\xi^{\rm KN}_e}N_e=-n_\gamma\sigma^e_{_{\rm KN}}N_e.
\label{prob1}
\end{eqnarray}
This rate equation is recast as
\begin{eqnarray}
\frac{d N_e(T_\gamma,\gamma_e)}{d\gamma_e} &=& -\frac{1}{\Delta\gamma^{\rm KN}_e(T_\gamma,\gamma_e)}N_e(T_\gamma,\gamma_e),
\label{prob2}
\end{eqnarray}
by using Eqs.~(\ref{accp1}) and (\ref{scat}). 

\begin{figure}
\centering
\includegraphics[width=3.0in]{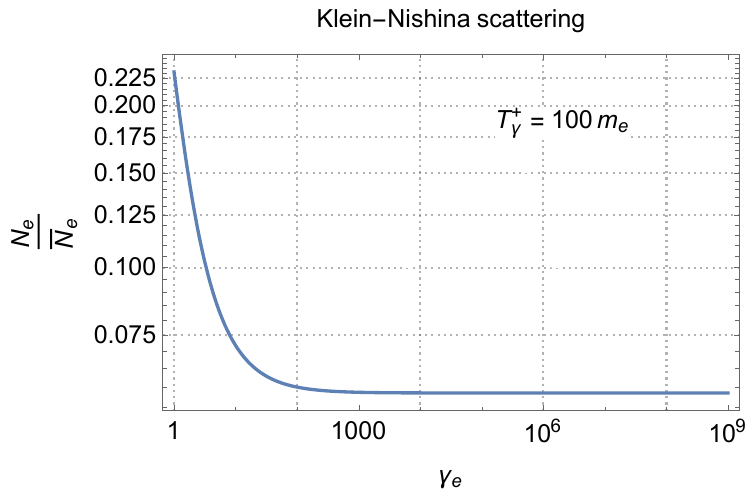}
\includegraphics[width=3.0in]{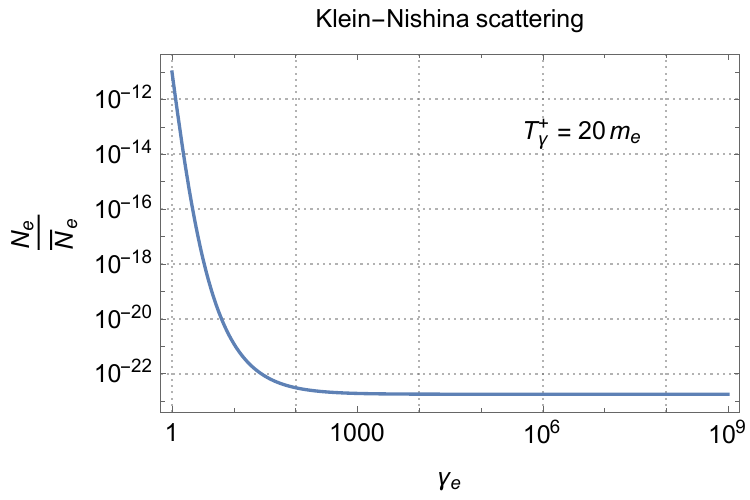}
\caption{We show the fraction $N_e/\bar N_e$ (\ref{prob2i}) of accelerated electrons drifting out of bulk electrons at different temperatures 
$T^+_\gamma\gg m_e$ near the horizon. 
The $N_e/\bar N_e$ value is exponentially sensitive to the horizon temperature $T_\gamma^+$. The probability of accelerated electrons reaching high energies 
is small for temperatures $T^+_\gamma \lesssim 10 m_e$, and the Compton-rocket effect and runaway dynamics might not be observationally relevant. These figures also represent the proton case by substituting the subscript 
$e\rightarrow p$.
}\label{notherm}
\end{figure}

The integration of the rate equation (\ref{prob2}) leads to the fraction (probability) of drifting electrons $N_e(T_\gamma,\gamma_e>1)$ out of bulk electrons $\bar N_e(T_\gamma,\gamma_e=1)$,
\begin{eqnarray}
\frac{N_e}{\bar
{N_e}}\equiv \frac{N_e(T_\gamma,\gamma_e>1)}{\bar  N_e(T_\gamma,\gamma_e=1)} &\approx & \exp -\int_1^{\gamma_e} \frac{1}{\Delta\gamma^{\rm KN}_e(T^+_\gamma,\gamma'_e)} {d\gamma'_e}.
\label{prob2i}
\end{eqnarray}
Here, we approximate the radial-dependent temperature 
$T_\gamma(r)$ by the characteristic temperature $T^+_\gamma$ for the same arguments 
discussed in Eqs.~(\ref{freet}) and (\ref{prob2i0}).  
Figure  \ref{notherm} gives the numerical integration (\ref{prob2i}) that shows the non-trivial fraction $N_e/\bar N_e$ 
of drifting electrons at ultra-high energies, which is in distinct contrast with the completely negligible 
one (\ref{prob2i0}) in the Thomson scattering case. This non-trivial number $N_e$ of drifting electrons ($\gamma_e\gg 1$) means a significant 
superthermal spectrum in the high-energy tail, which could be relevant for observations of UHE particles.

For different temperatures $T^+_\gamma\approx 100m_e\approx 50$ MeV and $T^+_\gamma\approx 20m_e\approx 10$ MeV, 
Figure \ref{notherm} shows the following spectral features of the probability that electrons get accelerated to a large Lorentz $\gamma_e$ factor:  
\begin{enumerate}[(a)]
\item for Lorentz factors $\gamma_e < 10^3$, the accelerated electron fraction $N_e/\bar N_e$ decreases as $\gamma_e$ increases, which may be approximated as a decay power law 
\begin{eqnarray}
\frac{N_e}{\bar N_e}\propto (\gamma_e)^{-\nu},~~ \nu>0, ~~ (\gamma_e < 10^3); \label{nngap}
\end{eqnarray}
\item for  Lorentz factors $\gamma_e> 10^3$, the fraction approach a constancy 
\begin{eqnarray}
\frac{N_e}{\bar N_e}~\propto (\gamma_e)^{0},~~~(\gamma_e> 10^3).\label{nnga}
\end{eqnarray}
\end{enumerate}
The feature (a) is expected in a usual physical situation where the probability of drifting electrons running to high energies becomes smaller, due to their kinetic energy dissipations in collisions with photons and pairs.
However, the feature (b) is peculiar, showing that the probability $N_e/\bar N_e$ does not decrease as (a), instead, becomes 
very flat when $\gamma_e>10^3$. 
These are the exact attributes of unstable runaway behaviour. Since the KN cross-section decreases as the drifting electrons' energies increase, they continue accelerating to higher and higher energies by the Compton-rocket effect without dissipating much energy into photons and pairs in the radiation fluid. 
Such an avalanche dynamics is crucially important
for producing UHE particles of Lorentz factors $\gamma_e\gg 1$. 

As the temperature $T_\gamma$ decreases and approaches $2m_e$, the asymptotic value $N_e/\bar N_e$ (\ref{nnga}) is exponentially suppressed by many orders of magnitude. Some technical difficulties occur in making numerical calculations. Therefore, using $\Delta \gamma^{\rm KN}_e$ (\ref{dgamma}) and $(N_e/\bar N_e)$ (\ref{prob2i}), we approximately calculate the asymptotic constancy (\ref{nnga}) for large $\gamma_e > 10^3$ 
\begin{eqnarray}
\left(\frac{N_e}{\bar N_e}\right)_{\gamma_e\gg 1} &\approx & \left(\frac{4T^+_\gamma}{m_e}\right)^\chi e^{-\chi},\quad \chi= \frac{m^{\rm eff}_e}{m_e}\left(\frac{m_e}{T^+_\gamma}\right)^2,
\label{const}
\end{eqnarray}
which is exponentially suppressed by small horizon temperatures $T^+_\gamma/m_e$, as can be seen in Figs.~\ref{therm500} and \ref{notherm}. This result (\ref{const}) implies the following features. (i) The significantly nontrivial fraction $N_e/\bar N_e$ of high-energy electrons ($\gamma_e\gg 1$) can only be produced in opaque fireshells or spots of the radiation fluid of photons and pairs at temperatures $T_\gamma > 10 m_e$, which can be present in highly energetic arenas of astrophysical sources and the early Universe.
(ii) In the case of the Sun ($T_\odot/m_e\ll 1$), the excess of large $\gamma_e$ electrons $N_e$ in the superthermal tail deviating from the blackbody radiation spectrum is too small to be relevant for observations. 
(iii) However, the probability of $\gamma_e\gg 1$ electrons' excess may be relevant for the detections 
in future laser experiments \cite{Exp_Sarri2025}
for the physics of strong radiation field intensities $I_\gamma >E^2_c/(8\pi)$, 
equivalently to high temperature $T^+_\gamma > 10 m_e$.

To close this section, we mention that for the optically thin system of electron-proton plasma interacting with photons at energies $\epsilon_\gamma\sim m_e$, Equations of types (\ref{prob10},\ref{prob20}) or (\ref{prob1},\ref{prob2}) 
were used in the numerical studies \cite{Faure2024}. Their results show the superthermal spectrum tail of runaway electrons $N_e$ drifting backwards to massive protons due to the electric field induced by the charge separation between electrons and protons. Here, instead, we study electrons and protons that are fully ionised and immersed inside an opaque radiation fluid of photons and pairs. They interact with photons and pairs at very high temperatures $T_\gamma\gg m_e$, energy density $\rho_\gamma\gg m_e^4$ and number density $n_\gamma\gg m_e^3$.

\section{Runaway protons pulled by runaway electrons}

\subsection{Electric field between drifting electrons and protons}

A non-trivial amount of high-energy electrons running away indicates that we cannot ignore protons in the dynamics of the Compton-rocket effect on electrons, because of the 
electric field $E_{pe}$ developed between separated electrons and protons. This is due to the very different cross sections $\sigma_{_T}\gg \sigma^p_{_T}\approx (m_e/m_p)^2\sigma_{_T}$ of photons scattering with electrons and protons. In the one-dimensional model, the Compton-rocket effect on protons
($p_p=m_p\beta_p\gamma_p$)  
\begin{eqnarray}
\frac {d\epsilon_p}{d\tau} &=&-\sigma^p_{_T}F_\gamma(r)[(4/3) (\beta_p \gamma_p)^2],\label{accpe}\\
\frac {d p_p}{d\tau} &=& \sigma^p_{_T} F_\gamma(r)[(2/3)(\beta_p \gamma_p)^2+1],
\label{accpp}
\end{eqnarray} 
is much smaller than the one (\ref{accp0}) for electrons.

As a result, charge separation and the electric field $E_{pe}$ between electrons and protons occur. It must back-react on the one-dimensional acceleration equations 
for both electrons (\ref{accp0}) and protons (\ref{accpp}), 
\begin{eqnarray}
\frac {d p_e}{d\tau} &=& \sigma_{_T} F_\gamma(r)[(2/3)(\beta_e \gamma_e)^2+1]-eE_{pe}> 0,\label{ce}\\
\frac {d p_p}{d\tau} &=& \sigma^p_{_T} F_\gamma(r)[(2/3)(\beta_p \gamma_p)^2+1]+eE_{pe}\approx eE_{pe}>0.\label{cp}
\end{eqnarray}
It is a highly non-linear dynamical system. 
We speculate the dynamical condition when photons and pairs' radiation force 
$\sigma_{_T} F_\gamma\propto T_\gamma^4(\beta_e \gamma_e)^2$ in Eq.~(\ref{ce}) is 
so strong that 
\begin{enumerate}[(i)]
\item a significantly non-trivial fraction $N_e/\bar N_e$ of accelerated electrons ($\gamma_e\gg 1$) drifting outward, i.e., $dp_e/d\tau >0$ in Eq.~(\ref{ce}) for electrons;
\item  it thus 
induces a strong and local electric field $E_{pe}=eN_e/(4\pi r^2)$, which is an electric field between a thin spherical capacitor of electron and proton layers in the one-dimensional model; 
\item this electric field binds electron and positron together, pulls massive protons $d p_p/d\tau\approx eE_{pe}>0$ in Eq.~(\ref{cp}), where the radiation force
term $\propto \sigma^p_{_T}$ in its right-handed side is negligible. As a result, a significantly non-trivial fraction $N_p/\bar N_p$ of accelerated protons ($\gamma_p\gg 1$) drift outwards in the radial direction, following the accelerated electrons' fraction $N_e/\bar N_e$.
\end{enumerate}

\subsection{Strong radiation field accelerating electrons and protons}

The bulk and drifting electron numbers are identical to the bulk and drifting proton numbers  
\begin{eqnarray}
\bar N_e = \bar N_p, ~~~ N_e = N_p, ~~~  N_e/\bar N_e=N_p/\bar N_p,
\label{neu}
\end{eqnarray}
because of the global charge neutrality of ionised electrons and protons, which are immersed inside an electrically neutral radiation fluid of photons and pairs.
The combination of Eqs.~(\ref{ce}) and (\ref{cp}) yields the 
one-dimensional acceleration equation for electrons and protons together  
\begin{eqnarray}
\frac {d (p_e+p_p)}{d\tau} &\approx& \sigma_{_T} F_\gamma(r)[(2/3)(\beta_e \gamma_e)^2+1],\label{accf0}
\end{eqnarray}
neglecting the radiation force on protons. 
Under the condition of a strong enough radiation flux $F_\gamma$ and the electric field $E_{pe}$ between electrons and protons, we assume that drifting electrons $N_e$ and protons $N_p$ are moving together
\begin{eqnarray}
\beta_e\approx \beta_p ~~~ {\rm and} ~~~  \gamma_e\approx \gamma_p,\label{epsame}
\end{eqnarray}
through the radiation fluid of photons and pairs. The acceleration equation (\ref{accf0}) is simplified as
\begin{eqnarray}
(m_e+m_p)\frac {d (\beta_e\gamma_e)}{d\tau} &\approx& \sigma_{_T} F_\gamma(r)[(2/3)(\beta_e \gamma_e)^2+1].\label{accf}
\end{eqnarray}
This is the same as the electron acceleration 
equation (\ref{accp0}) or (\ref{accp1}) by introducing an effective electron mass
\begin{eqnarray}
m^{\rm eff}_e=m_p+m_e\approx m_p,
\label{effm}
\end{eqnarray}
which previously appeared in Eq.~(\ref{accp1}) without explanations.
Similarly, the combination of (\ref{acce0}) and (\ref{accpe}) yields the energy-gain equation for accelerated electrons and protons
\begin{eqnarray}
\frac {d\epsilon_{ep}}{d\tau} &\approx &-\sigma_{_T}F_\gamma(r)[(4/3) (\beta_e \gamma_e)^2],~~~ \epsilon_{ep}=\epsilon_e+\epsilon_p,\label{accfp}
\end{eqnarray} 
where $\epsilon_e=m_e\gamma_e$ and $\epsilon_p=m_p\gamma_p$.

Equations (\ref{accf}) and (\ref{accfp}) qualitatively describe an accelerating system of electrons and protons driven by the radiation force, assuming electrons and protons are tightly associated by their own electric field 
and accelerated to the approximately same Lorentz factor $\gamma_e\approx \gamma_p$. In this approximation, all dynamical equations remain the same as those for electrons, except for replacing the electron mass $m_e$ with the effective electron 
mass $m^{\rm eff}_e=m_p+m_e$ (\ref{effm}) for a heavy system of electrons and protons. 
As previously anticipated, the effective mass $m^{\rm eff}_e \gg m_e$ replaces $m_e$ in the dynamic equations (\ref{accp1}) and (\ref{accf}). 
As a consequence, the large temperature $T_\gamma > 10 m_e$ radiation flux $F_\gamma$ acting on an electron is necessarily 
required to accelerate a heavy system of both electrons and protons.

Such a dynamical configuration should be realised in an opaque radiation fluid of photons and pairs at large temperatures $T_\gamma\gg m_e$ and number $N_\gamma \gg \bar N_e=\bar N_p$, where the bulk numbers $\bar N_e$ and $\bar N_p$ of fully ionised electrons and protons are small contaminations.  
Otherwise, if the radiation fluid temperature $T_\gamma$ is not large enough, instead of accelerating protons, the bulk electrons $\bar N_e$ are accelerated backwards to massive protons by the induced electric field  $E_{pe}$, as numerically studied in Ref.~\cite{Faure2024}.

In addition, when electrons and protons drift through the radiation fluid of photons and pairs, they undergo Klein-Nishina scattering processes. The KN cross section $\sigma^p_{_{\rm KN}}$ of drifting protons 
is much smaller than the one $\sigma^e_{_{\rm KN}}$ (\ref{knh}) of drifting electrons, i.e., $\sigma^p_{_{\rm KN}}\sim {\mathcal O}[(m_p/m_e)^2]\sigma^e_{_{\rm KN}}$. 
Therefore, we neglect the cross section $\sigma^p_{_{\rm KN}}$ of drifting protons in the Klein-Nishina scattering processes between an electron-proton pair and the radiation fluid of photons and pairs.
In these approximations, all analyses and results in Section \ref{run} remain the same. 
Both drifting electrons and protons run away, becoming UHE particles. 
Figures \ref{notherm} also represent the fraction of accelerated protons with the substitution $e\rightarrow p$. 
In the following, without further specification, all discussions and calculations apply to a bound system of electrons and protons with an effective electron mass $m^{\rm eff}_e=m_p+m_e$. Note that for the case $\gamma_e\approx \gamma_p$, the accelerated protons' energy $\gamma_pm_p$ is ${\mathcal O}[(m_p/m_e)]$  larger than the accelerated electrons' energy $\gamma_em_e$.

\section{Luminosities of UHE emission and blackbody radiation}\label{lumep}

The accelerated electrons $N_e$ and protons $N_p$ 
reach ultra-high energies and stream out of the fireshell, carrying away the thermal energy $E_\gamma$ (\ref{totale}) of the radiation fluid of photons and pairs. 
We will calculate the luminosity of ultra-high-energy emissions of electrons and protons, as well as the time variation of the fireshell thermal energy $E_\gamma$. 
As will be calculated below, the luminosity and the rate $\dot E_\gamma/E_\gamma$ of thermal energy dissipation are small. Up to a certain timescale, the processes of thermal energy transfer and temperature cooling are nearly adiabatic, and the fireshell thermal equilibrium is approximately maintained with slowly decreasing temperature. Therefore, we regard the fireshell as an approximate steady energy $E_\gamma$ reservoir, like the Sun. As a result of thermal energy dissipation and strong gravitational effect, the fireshell eventually cools down and shrinks in size.

\subsection{Blackbody radiation luminosity and MeV annihilation line} 

Due to the large opacity and temperature gradient in the centre of the fireshell ($r_+<r<2r_+$), internal photons diffuse 
outwards in the radial direction. 
At the fireshell outer boundary $r=2r_+$, where the temperature is about $2m_e$ and pair annihilation occurs, it becomes optically thin, and photons stream out as a continuous blackbody radiation of the characteristic temperature 
$2m_e$. One obtains the blackbody luminosity by solving the radiation transfer (photon diffusion) equation \cite{böhm1989introduction}, %\cite{Ruffini2000a}, 
\begin{eqnarray}
L^0_\gamma &\sim&  4\pi (2r_+)^2 \sigma_{_{\rm SB}}d_f(T_\gamma^+)^4c\left(\frac{\xi_\gamma}{2r_+}\right)\approx 4\left(\frac{\xi_\gamma}{2r_+}\right)L^+_\gamma.
%\approx 4\pi^3 d_f 10^{-22}\left(\frac{M}{M_\odot}\right)^2 \left(\frac{M_\odot c^2}{10^{-5}{\rm sec}}\right).
\label{lumig}
\end{eqnarray}
The continuous radiation time scale is
\begin{eqnarray}
\frac{E_\gamma}{L^0_\gamma} &\approx & \left(\frac{1}{4\eta}\right)\left(\frac{\delta\ell}{c}\right)\left(\frac{2r_+}{\xi_\gamma}\right),
%\approx 4\pi^3 d_f 10^{-22}\left(\frac{M}{M_\odot}\right)^2 \left(\frac{M_\odot c^2}{10^{-5}{\rm sec}}\right).
\label{lumigt}
\end{eqnarray}
where the photon luminosity $L^+_\gamma=\eta L_{\rm pl}$ (\ref{bconf}) and mean-free path $\xi_\gamma\sim 10^{-4}\lambda_e$ 
(Fig.~\ref{plasma}) are inside the centre of the fireshell of photons and pairs.  
In this case, the screen factor $\xi_\gamma/2r_+\sim 8.3\times 10^{-22}$ is small for the fireshell size $\sim 2r_+$ (\ref{bhcon}). 
The blackbody radiation luminosity (\ref{lumig}) is a constant $L^0_\gamma\sim 10^{-22}L_{\rm pl}$, and the time scale (\ref{lumigt}) is large. These values depend on the fireshell parameters (\ref{bhcon}), (\ref{bcon}),  
(\ref{bcont}), and (\ref{bconf}).

This emission is analogous to the blackbody radiation from the Sun's surface. 
To compare and contrast, we mention the Sun's surface halo of the temperature $\sim 6\times 10^3C^\circ\approx 0.55$ eV 
and the Sun radius $6.96\times 10^{10}$cm, where electrons are not relativistic ($\langle\gamma_e\rangle\approx 1$) and their densities are much smaller than the fireshell under discussion. 
The Sun's luminosity is $L_\odot= 3.83\times 10^{33} {\rm ergs/sec}\approx 10^{-27} L_{\rm pl}$.
Thermal photons diffuse from the Sun's core at the temperature $T_\odot=15\times 10^6 C^0=1.5\times 10^4 {\rm eV} =3\times 10^{-2} m_e$. While in the fireshell case, the blackbody 
has a characteristic temperature $\sim 2 m_e$. Moreover, the radiation spectrum contains the electron-positron annihilation line and reheating of temperature increment $(11/4)^{1/3}$ before 
and after annihilation around $2 m_e \approx 1$ MeV \cite{Ruffini1999}.  

\begin{figure}
\centering
\includegraphics[width=3.2in,height=2.2in]{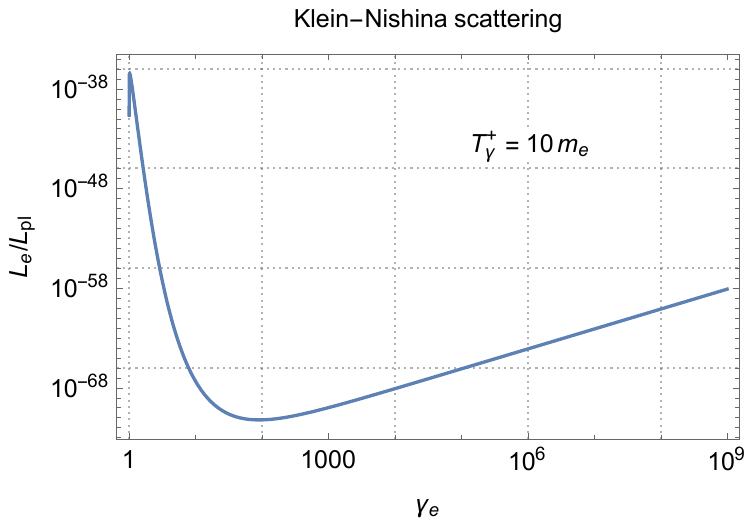}%height
\includegraphics[width=3.2in,height=2.2in]{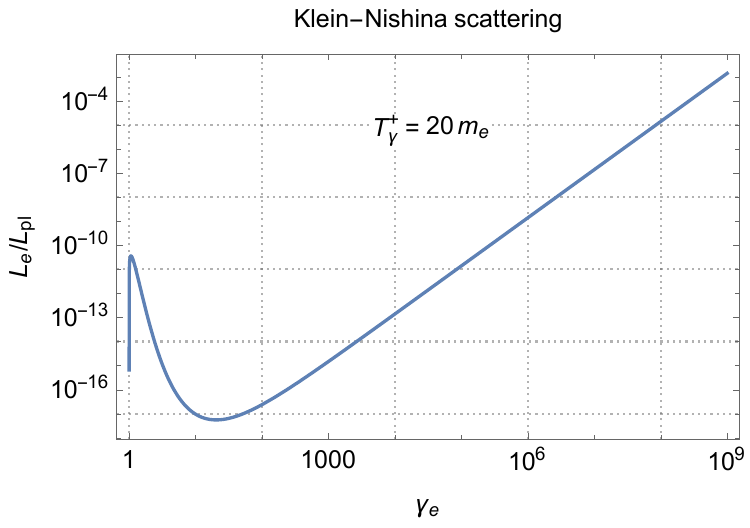}%height
\caption{For the purpose of illustrating the spectral behaviour of UHE particle luminosity, we select the parameters: the horizon temperature (\ref{bcont}) $T_\gamma^+=10m_e,20m_e$ 
and acceleration number $A=10^{-3}$ (\ref{lumA}), to plot high-energy electron and proton luminosities $L_e/L_{\rm pl}$ (\ref{lumie0}) in terms of the particle Lorentz factor $\gamma_e$.}\label{Legamma}
\end{figure}

\subsection{UHE emission luminosity and energy spectrum}

Section \ref{run} shows that the KN cross section (\ref{knh}) decreases as accelerated electron and proton energies increase. Therefore, we expect high-energy electrons to stream
out and carry away the fireshell energy $E_\gamma$ of photons and pairs. 
Using (i) the rates of radiation energy transfer to an accelerated 
electron and proton pair (\ref{accfp}), (ii) total bulk electron number $\bar N_e=B N_\gamma$ (\ref{bulkb}) 
and (iii) photon-pair number $N_\gamma$ (\ref{totaln}), we obtain the luminosity of high-energy electrons (protons) of the Lorentz factor $\gamma_e\approx \gamma_p$,
\begin{eqnarray}
L_e&=&\frac {dE_e}{dt} \approx N_e\sigma_{_T}F_\gamma(r_+)[(4/3) (\beta_e \gamma_e)^2]\nonumber\\
&=&A\left(\frac{N_e}{\bar N_e}\right)\left(\frac{n^+_\gamma}{m^3_e}\right)(4/3)(\beta_e \gamma_e)^2L_\gamma^+,\label{lumie0}
%\\ &\approx &8.0\times 10^{37 }B\left(\frac{N_e}{\bar N_e}\right)\left(\frac{n^+_\gamma}{m^3_e}\right)(\beta_e \gamma_e)^2m_e^2, \label{lumie}
\end{eqnarray}
where the ``acceleration number'' is defined as
\begin{eqnarray}
A=B(\sigma_{_T}\delta\ell m_e^3),
\label{lumA}
\end{eqnarray}
indicating the effective number of bulk electrons $\bar N_e$ accelerated inside the volume $(\sigma_{_T}\delta\ell)$ of one-dimensional tube of photon-electron interactions. We treat the $A$ as a parameter, depending on the baryon matter contamination $B\ll 1$ and the firespot size $\delta\ell\gg \lambda_e$, the quantity  $(\sigma_{_T}\delta\ell m_e^3)= (8\pi/3)\alpha^2(\delta\ell/\lambda_e)$ can be much larger than one. 

We point out that the obtained UHE emission luminosity (\ref{lumie0}) exhibits three distinct features: 
\begin{enumerate}[(1)]    
\item The luminosity sensitively depends on the horizon  temperature $T_\gamma^+$ (\ref{bcont}), via 
the photons and pairs number density $n^+_\gamma\equiv n_\gamma(T_\gamma^+)$ (\ref{tem}) and the fraction $(N_e/\bar N_e)$ (\ref{prob2i}) of electrons accelerated up to the Lorentz factor $\gamma_e$. 
Figure \ref{notherm} shows that the fraction $N_e/\bar N_e$ is exponentially suppressed by small temperatures 
$T^+_\gamma \lesssim 10 m_e$. Therefore, the UHE emission luminosity (\ref{lumie0}) is completely negligible in these cases, unless the non-trivial acceleration number $A\gg 1$ (\ref{lumA}), which relates to the baryon-loading $B$ and size $\delta\ell$ of the fireshells (spots) of energetic photons and pairs.
\item    
The fraction $N_e/\bar N_e$ properties (\ref{nngap}), (\ref{nnga}) and Figure \ref{notherm} imply the UHE emission luminosity (\ref{lumie0}) of particles' energies 
$m_e^{\rm eff}\gamma_e$ follows the spectral behaviour
\begin{eqnarray}
(i)~ L_e\propto (\gamma_e)^{2-\nu},~(\gamma_e < 10^3); ~~~(ii)~ L_e~\propto (\gamma_e)^{2},~(\gamma_e> 10^3),\label{spec}
\end{eqnarray}
as shown in Fig.~\ref{Legamma}. This is consistent with the Compton-rocket effect (\ref{acce0}) and (\ref{accp0}) \cite{ODell1981}, showing that relativistic electrons gain more energy ($\gamma_e> 1$) from radiation photons than non-relativistic ones ($\gamma_e\approx 1$).
\item Similar to the blackbody radiation (\ref{lumig}), the UHE emission is not a transient phenomenon, but rather a lasting, continuous radiation. The reason is that the fraction $(N_e/\bar N_e)$ becomes tiny,  and the emission time scale 
\begin{eqnarray}
\frac{E_\gamma}{L_e} &\approx & \left(\frac{\delta\ell}{c}\right)\left[A\left(\frac{N_e}{\bar N_e}\right)\left(\frac{n_\gamma^+}{m^3_e}\right)(4/3)(\beta_e\gamma_e)^2\eta\right]^{-1},
%\approx 4\pi^3 d_f 10^{-22}\left(\frac{M}{M_\odot}\right)^2 \left(\frac{M_\odot c^2}{10^{-5}{\rm sec}}\right).
\label{lumiet}
\end{eqnarray}
becomes even larger than the blackbody one (\ref{lumigt}).
\end{enumerate}
In the second point (2), the luminosity increases with the UHE particles' energy $m_e^{\rm eff}\gamma_e$, but the increment in luminosity $L_e\propto (\gamma_e)^{2}$ with $\gamma_e$ does not run to infinity $L_e\propto (\gamma_e)^{2}\rightarrow \infty$ in time. The reason is due to the back reaction that the large luminosity $L_e$ of large $\gamma_e$ decreases the fireshell temperature $T_\gamma^+$ and energy $E_\gamma$ fast in time, as required by total energy conservation. 
As a result, the accelerated electron fraction $N_e/\bar N_e$ (\ref{const}) and the UHE emission luminosity $L_e$ (\ref{lumie0}) of large $\gamma_e$ decrease 
fast in time, as will be shown below. 

To compare two emission processes from the fireshell, 
we calculate the ratio of the blackbody radiation (\ref{lumig}) and UHE emission (\ref{lumie0}) luminosities given by
\begin{eqnarray}
\frac{L_e}{L^0_\gamma}&=& A\left(\frac{N_e}{\bar N_e}\right)\left(\frac{n^+_\gamma}{m^3_e}\right) (\beta_e \gamma_e)^2\left(\frac{\xi_\gamma}{2r_+}\right)^{-1}.\label{lumir}
\end{eqnarray}
There are two possibilities:
\begin{enumerate}[(i)]
\item 
the blackbody radiation is dominant $L_e/L^0_\gamma \ll 1$ for low temperature $T^+_\gamma \gtrsim m_e$ and small fraction $N_e/\bar N_e\ll 1$;  
\item 
the UHE emission is dominant $L_e/L^0_\gamma \gg 1$ for large temperature $T^+_\gamma \gg m_e$ and the significantly non-trivial fraction $N_e/\bar N_e$. 
\end{enumerate}

\noindent The former case (i) is well studied, and the blackbody luminosity is small due to the fireshell's large opacity $(r_+/\xi_\gamma)\gg 1$. We are primarily interested in the latter case (ii) for $\gamma_e \gg 1$, when $T^+_\gamma \gg m_e$ and the UHE emission dominantly dissipates the energy $E_\gamma$ (\ref{totale}) of the fireshell. 

\begin{figure}
\centering
\includegraphics[width=3.2in,height=2.2in]{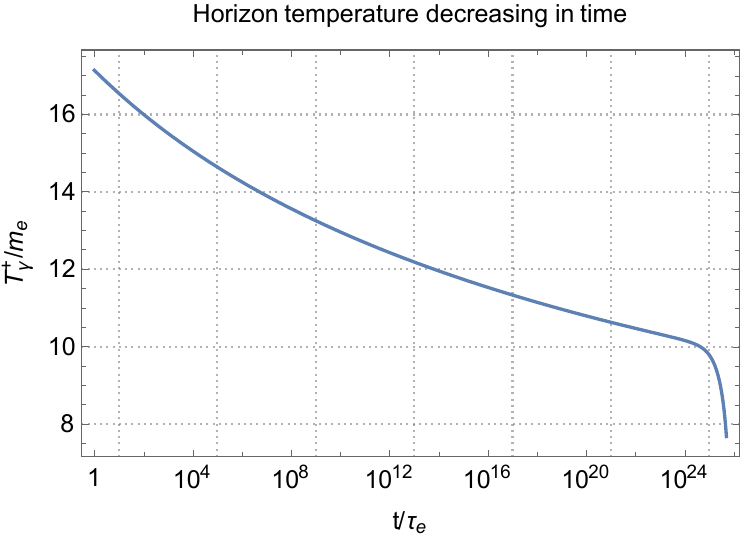}%height
\includegraphics[width=3.2in,height=2.2in]{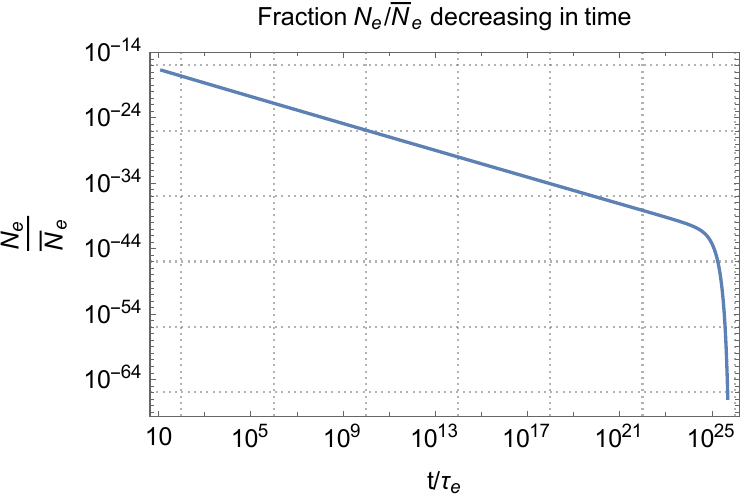}\\%height
\includegraphics[width=3.2in,height=2.2in]{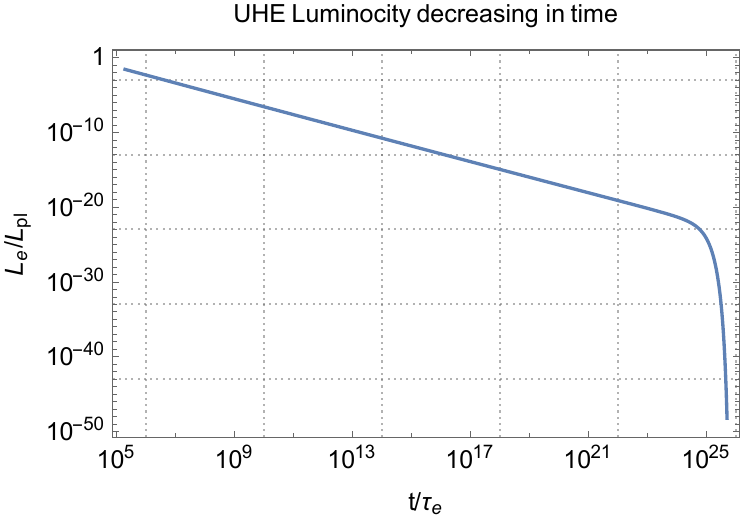}%height
\includegraphics[width=3.2in,height=2.2in]{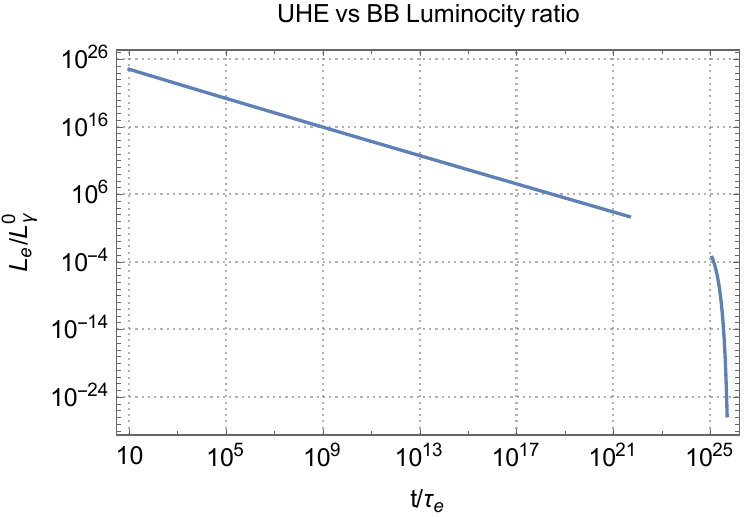}%height
\caption{For illustrating the numerical results, we adopt the initial horizon temperature $T_\gamma^+(0)=100m_e\approx 50$ MeV (\ref{bcont}) and the acceleration number $A=10^{-3}$ (\ref{lumA}). 
We consider the Lorentz factor $\gamma_e\sim 10^9$  channel, corresponding to UHE electron energies $\sim 10^{3}$ TeV and proton energies $\sim 10^{3}$ PeV. 
The Compton time $\tau_e=(\lambda_e/c)=1.28\times 10^{-21} {\rm sec}$, and
the final time $10^{26}\tau_e$ is about days. The $L_e/L_\gamma^0$ broken line (log-log plot) indicates the transition tuning place from the UEH emission dominant $L_e/L^0_\gamma \gg 1$ to the blackbody (BB) emission dominant $L_e/L^0_\gamma \ll 1$. For different values of $\gamma_e$ and $T_\gamma^+(0)$, the results will exhibit the same behaviours in time variations. The UHE emission luminosity $L_e$ increases for large initial temperature $T_\gamma^+(0)$ and acceleration number $A$. The emission duration lasts longer for smaller $\gamma_e$ channels of UHE particles.
}\label{Letime}
\end{figure}

\subsection{Fireshell temperature and UHE emission decrease in time}

The blackbody emission (\ref{lumig}) and the UHE emission (\ref{lumie0}) dissipate the fireshell thermal energy $E_\gamma$ (\ref{totale}), therefore, the total energy conservation yields   
\begin{eqnarray}
-\frac{dE_\gamma}{dt}= \frac{dE^0_\gamma}{dt}+\frac{dE_e}{dt}=L^0_\gamma+L_e.\label{dedt}
\end{eqnarray}
We use this equation together with Eqs.~(\ref{totale}), (\ref{lumie0}) and (\ref{lumir})
to determine how the horizon temperature $T^+_\gamma(t)$ decreases in time,
\begin{eqnarray}
\frac{dT^+_\gamma(t)}{dt}&=& -1.23B\left(\frac{N_e}{\bar N_e}\right) \left(\frac{\sigma_{_T}}{4\pi r_+^2}\right)(\beta_e \gamma_e)^2L_\gamma^+\left(1+\frac{L_\gamma^0}{L_e}\right)\nonumber\\
&=&-10.4 \eta A\left(\frac{N_e}{\bar N_e}\right)(\beta_e\gamma_e)^2m_e^2\left(1+\frac{L_\gamma^0}{L_e}\right),\label{dedt1}
\end{eqnarray}
where the factor $\eta$ in the second line comes 
from the boundary condition (\ref{bhcon}). In the fireshell decay equation (\ref{dedt1}), we particularly consider a large $\gamma_e\gg 1$ channel, 
because it dominates the energy dissipation of the fireshell. Before presenting numerical calculations, we note that the gravitational redshift factor in energy and the gravitational dilation in time are simplified in the luminosities $L_\gamma^0$ (\ref{lumig}) and $L_e$ (\ref{lumie0}),
and the temperature variation $dT^+_\gamma/dt$ (\ref{dedt1}).

For a given channel $\gamma_e\gg 1$ value and an initial value $T^+_\gamma(0)$ at the time $t=0$, we use the numerical result $N_e/\bar N_e$ (\ref{prob2i}) 
as a function of $T^+_\gamma(t)$ to integrate 
the fireshell decay equation (\ref{dedt1}) over the time. Furthermore, substituting the numerical solution $T^+_\gamma(t)/m_e$ into $(N_e/\bar N_e)$ and $(n_\gamma^+/m_e^3)$ in Eq.~(\ref{lumie0}), we calculate the UHE emission luminosity $L_e(t)$ declining as a function of the time $t/\tau_e$. We calculate these integrations numerically by using the Mathematica program.
Figure \ref{Letime} shows that the numerical results 
of the horizon temperature $T^+_\gamma(t)/m_e$, the accelerated electron fraction $N_e(t)/\bar N_e$, and the UHE emission luminosity $L_e(t)$ monotonically decrease in time $t/\tau_e$ varying up to $10^{26}\tau_e$, which is about days. The results show two phases:
\begin{enumerate}[(i)]
\item 
The initial phase:
the UHE emission initially dominates over the blackbody radiation $L_e/L_\gamma^0 > 1$, and the turning point to $L_e/L_\gamma^0 < 1$ occurs 
around $t=10^{25}\tau_e$ when the temperature $T^+_\gamma\approx 10 m_e$, and the accelerated electron fraction $N_e(t)/\bar N_e$ decreases significantly in time.  
\item 
The latter phase: the fireshell cooling process is attributed to the blackbody radiation $L_\gamma^0$ (\ref{lumig}) 
and the temperature $T^+_\gamma(t)$ decreases in time at a different slope (power law) from the one in the initial phase. 
\end{enumerate}
\noindent

In the UHE emission luminosity-dominant phase 
$L_e/L^0_\gamma \gg 1$, we neglect the small time-variation effect of the blackbody luminosity 
$L_\gamma^0(t)$ (\ref{lumig}) caused by the time varying mean-free path $\xi_\gamma=(\sigma_\gamma n^+_\gamma)^{-1}$ (\ref{ttime}) and number density $n^+_\gamma\propto [T^+_\gamma(t)/m_e]^3$ of photons and pairs in the fireshell.
The effect should probably be considered in the blackbody luminosity $L_e/L^0_\gamma \ll 1$ dominant phase, when the horizon temperature $T^+_\gamma < 10 m_e$, electron-positron pair annihilation into two photons occurs, and the fireshell opacity decreases. 
As a consequence, the TOV equilibrium (\ref{tov0}) is 
gradually lost and the fireshell (or fire spot) shrinks and disappears. These are not the topics of this study and will be subjects of future investigations. 

\section{Characteristics relevant for observations and experiments}\label{obs}

\subsection{Peculiar features of UHE particles produced in such a scenario}
 
We highlight some qualitative characteristics of UHE particles produced by the scenario discussed in previous sections:  
\begin{enumerate}[(1)]
\item The mechanism occurs at local microscopic length and time scales, 
over which the photon-pair fluid temperature ($T_\gamma \gg  m_e$) has a large gradient, resulting in a strong radiation force for electron and proton acceleration. Therefore, the mechanism does not essentially depend on the global symmetry, macroscopic length and time scales of high-energy astrophysical environments and strong gravitational processes, 
which create such a fire spot of highly dense and energetic fluid of photons and pairs. These fire spots contain thermal energy $E_\gamma$ (\ref{totale}) of the radiation fluid of photons and pairs, which are
energetic sources for UHE particle productions.
Therefore, the trapped fireshell (halo) near the horizon is only an example to simplify calculations and illustrations. The fire spots
can be formed, for instance, in gravitational binary coalescence and accretion cases, in the early Universe, and may be observed in future laser experiments.  

\item The horizon temperature $T^+_\gamma > 10 m_e$, as if an effective overcritical classical 
electric field $E_{\rm eff}\sim {\mathcal O}(10)E_c$ provides a significant probability of accelerating electrons and protons to UHE, up to TeV, PeV, EeV, and even higher. The total luminosity of UHE particles increases as the acceleration number $A$ increases, namely, more electrons in the fireshell participate in the acceleration process. The UHE particle luminosity (\ref{lumie0}) could be significant for small temperatures $2m_e<T^+_\gamma < 10 m_e$ and large acceleration numbers $A$. However, we cannot show the results, because the numerical precision is limited in calculating the exponential factor $e^{-m_{\rm eff}/T_\gamma^+}$. 
 
\item The UHE emission is not a transient phenomenon, such as a burst, but a long-lasting continuous radiation. UHE particles have peculiar spectral behaviours in high and low energies: approximate power laws $\propto \gamma^{-\nu}_e$ 
for small Lorentz factor $\gamma_e \lesssim 10^3$, and $\propto \gamma_e^2$ for larger $\gamma_e$. The luminosity of
UHE particles should be suppressed for very large $\gamma_e$ due to the back reaction of UHE particle production on the sources of fire spots.
\item Two fire-spot parameters are essential for characterising different sources of producing 
UHE particles: 
\begin{eqnarray}
(T^+_\gamma,~A).
\label{ta}
\end{eqnarray}
The temperatures $T^+_\gamma$ of fire spots of photons and pairs
are crucial for UHE particle spectra ($\gamma_e$). 
The acceleration number $A$ of electrons (protons) participating in acceleration dynamics is relevant to the total luminosity of UHE particles. 
For the appropriate values 
$T^+_\gamma$ and $A$ of the fire spots of astrophysical sources, the resultant UHE emission luminosity (emission probability) 
and energy spectra (Lorentz factor $\gamma_e$ or $\gamma_p$) can be relevant for the observations of UHE particles. In general, astrophysical fire-spot sources with large (small) $T^+_\gamma$ and $A$ values produce large (small) luminosity and energy of UHE particles. While the sources with large $T^+_\gamma$ and $A$ 
values are less probable. This may explain why the observed UHE particle luminosity monotonically 
decreases with energy.
\end{enumerate}
In principle, the primary UHE electrons and protons are energetic enough to produce secondary VHE neutral photons and neutrinos through their decays and/or interactions with low-energy particles in the environments of sources of fire spots. 
Namely, UHE electrons and protons interact with low-energy fields and particles in the medium around their sources, producing VHE photons and neutrinos. 
Since these interactions are highly energetically 
forward scattering, resultant secondary
VHE particles carry the characteristics of primary UHE particles.  This implies that UHE and VHE particle events should be observationally related to each other and correlate with astrophysical sources.

\subsection{Possible relevance for observations and experiments}\label{vhegrb}

In this article, we focus only on revealing the novel scenario and mechanism of generating UHE particles, and 
pay no attention to quantitatively discussing this scenario compared with the observations of UHE 
and VHE particles. 
%The reasons are that the data are too few to quantify the UHE particle luminosity properties, which also depend on other factors, e.g., absorption, in addition to the mechanism of accelerating charged electrons and protons. 
Nevertheless, we mention some observationally and experimentally relevant features, in particular for UHE and VHE particles accompanying GRB phenomena observed. 

Interacting with magnetic fields, soft photons 
and other low-energy particles around sources, these UHE electrons and protons from the opaque radiation fluid spot of highly energetic photons and pairs produce 
VHE photons and neutrinos to TeV or PeV energy scales.  
We come to GRBs' cases. As shown in Fig.~\ref{Letime}, the trapped fireballs (horizon halos) cool via continuous emission of UHE electrons and protons. Interacting with the surrounding fields and materials, the former 
mainly leads to VHE photons, and the latter to VHE neutrinos.  
This scenario strongly implies the logical and causal reasoning features that are relevant to the observations of VHE photons accompanying GRBs' phenomena: 
\begin{enumerate}[(1)]
\item UHE electrons (protons) and the subsequent VHE photons and neutrinos are continuous emissions in the fireshell lifetime. They should correlate with each other with the GRBs' sources in observations. 
\item Before outward expanding fireballs become optically thin, we probably cannot see VHE photons because of great opacity due to two-photon annihilation $\gamma\gamma\rightarrow e^+e^-$  
of the Breit-Wheeler process (\ref{ggee}). During the prompt emissions for bursts, the VHE photons most probably observed are after expanding fireballs become transparent, namely, VHE photon luminosity increases soon after the main GRBs' bursts occur \footnote{The features (1) and (2) were pointed out in Ref.~\cite{Xue_2021}.} 
\item 
During the afterglow era, UHE protons $\gamma_p\gg 1$ emitted from the horizon fireshell continuously collide with the ejector of 
the Lorentz factor $\Gamma<\gamma_p$, which then emits VHE photons and neutrinos. 
Therefore, their (VHE) light curves follow the law of the ejector's kinematic motion in time, similarly to the external shock emission of the ejector colliding with the interstellar medium. However, their (VHE) energy spectra become 
softer in time because the horizon fireshell cool down. The time and luminosity of UHE and VHE continuous emissions depend on the horizon fireshell temperature $T_\gamma$ and acceleration number $A$ (\ref{ta}), as well as the target (ejector) density.
\end{enumerate}

\noindent In contrast, according to the usual modelling interpretation 
of the TeV emission in GRBs, ultra-relativistic electrons are accelerated by internal
shocks of fireballs during the prompt emission or by
external shocks between the ejector and interstellar medium (ISM) during the afterglow phase.

In addition to astrophysical sources, the scenario and mechanism in principle apply to the early Universe at large particle density and high temperature $T_\gamma > 2m_e$ before the Universe becomes transparent. 
Local dense and inhomogeneous radiation fields accelerate charged particles to high energies. 
It implies that in the cosmic background observed today, there should be imprints of excess UHE and VHE particles originating from the early Universe.  Moreover, such an energetic  
spot would be created in future by strong laser-beam experiments in ground laboratories. The probability of producing UHE particles could be non-trivial for detection, if the strong laser field spot
is not only sufficiently energetic $T_\gamma>10 m_e$ and dense $A\gg 1$, but also has a lifetime longer than the time scale $\tau^e_{_{\rm KN}}=(\sigma^e_{_{\rm KN}}n_\gamma c)^{-1}$ (\ref{kntime}) for the runaway process.

\section{Summary and remarks}

In this article, we study the possibility of how electrons and protons 
are accelerated to UHE particles in an opaque fire spot of dense photons and $e^-e^+$ pairs at a large temperature $T_\gamma\gg m_e$. 
We have revealed a possible acceleration mechanism for UHE charged particles, which mainly attribute to (i) the Compton-rocket effect of strong radiation force acting on electrons for acceleration; (ii) the instability and runaway dynamics occur when accelerated electrons undergo the KN scattering process, whose cross section decreases with increasing electron energy; (iii) such drifting electrons develop a local electric field acting on protons for acceleration.  
For qualitative analysis and theoretical illustrations, we use the Compton-rocket equation for electrons' energy gain
and the rate equation of electrons' energy loss in Thomson and 
KN scatterings in a simplified one-dimensional model.  As a result, we obtain the non-trivial probability of accelerating electrons and protons reaching a large Lorentz factor $\gamma_e\sim 10^9$ over a small distance $d\sim 10^{-2}$ cm.

As an astrophysical source for the opaque fire spots of dense photons and pairs at large temperatures, 
we consider the opaque fireballs formed in gravitational collapses. The outer parts of fireballs expand, accounting for the phenomena of GRBs. Whereas the inner parts of fireballs, fireshell, slowly contract and get dense near the horizon due to strong gravity and large pressure. The inner part of fireballs, fireshell as an energy source, possesses
large opacity and thermal energy with a small amount $B$ of electrons and protons from baryon matter contamination, which are comparable with counterparts of the outer part of fireballs. 
Approximating the fireshell as a metastable state, we calculate the luminosities and spectra of UHE emission and blackbody radiation. In this scenario, we point out distinct features of produced UHE and VHE particles, which are  relevant for observations and in contrast with the usual interpretations of the TeV emission in
GRBs.  

However, we emphasise that the systems and dynamics under investigation are rather complex, consisting of photons and pairs, protons and electrons, various interacting processes, whose times and lengths run from the microscopic to the macroscopic scales. In these preliminary studies,
the models and conditions used to present the idea and scenario are simplified, the fraction $N_e/\bar N_e$ and the analyses used to show the results are qualitative. The one-dimensional electric field between drifting protons and electrons should be generalised to the three-dimensional case. Full interaction rates and cross sections depending on energy and momentum should be considered to investigate how the instability of electron perturbations develops in the strong radiation force of dense and energetic photons and pairs, leading to an avalanche runaway process. 
Needless to say, it is necessary to develop a numerical code for performing full simulations to verify the present scenario of accelerating electrons and protons to ultra-high energies. The dynamics are attributed to the radiation force 
in an opaque fire spot of dense photons and pairs at temperatures that exceed the electron mass and have a large spatial gradient in a preferential direction. 
Moreover, it is also necessary to study how such firespots can be established in high-energy astrophysical processes in strong gravitational environments, and determine their temperatures, sizes and lifetimes. In observations, it is worthwhile to phenomenologically examine the three distinct features of VHE photon emissions in GRBs mentioned in Sec.~\ref{vhegrb}, 
and possible relations between observed cosmic ray events of UHE particles and VHE emissions of photons and neutrinos, and correlations
with high-energy astrophysical sources.

%recapitulating

%\bibliographystyle{JHEP}
%\bibliography{WangYu}

\begin{thebibliography}{10}

\bibitem{Globus2025a}
N.~Globus and R.D.~Blandford, \emph{Ultrahigh-energy cosmic rays}, \href{https://doi.org/10.1146/annurev-astro-052622-033150}{\emph{Annual Reviews of Astronomy and Astrophysics} {\bfseries 63} (2025) 339} [\href{https://arxiv.org/abs/2505.21846}{{\ttfamily 2505.21846}}].

\bibitem{LHAASO:2021gok}
Z.~Cao and others (LHAASO~collaboration), \emph{{Ultrahigh-energy photons up to 1.4 petaelectronvolts from 12 $\gamma$-ray Galactic sources}}, \href{https://doi.org/10.1038/s41586-021-03498-z}{\emph{Nature} {\bfseries 594} (2021) 33}.

\bibitem{Acciari2019}
{\scshape MAGIC} collaboration, \emph{Teraelectronvolt emission from the $\gamma$-ray burst GRB 190114c}, \href{https://doi.org/10.1038/s41586-019-1750-x}{\emph{Nature} {\bfseries 575} (2019) 455} [\href{https://arxiv.org/abs/2006.07249}{{\ttfamily 2006.07249}}].

\bibitem{IceCube:2023ame}
{\scshape IceCube} collaboration, \emph{{Observation of high-energy neutrinos from the Galactic plane}}, \href{https://doi.org/10.1126/science.adc9818}{\emph{Science} {\bfseries 380} (2023) 9818} [\href{https://arxiv.org/abs/2307.04427}{{\ttfamily 2307.04427}}].

\bibitem{Rees:1992ek}
M.J.~Rees and P.~Meszaros, \emph{{Relativistic fireballs - energy conversion and time - scales}}, {\emph{Mon. Not. Roy. Astron. Soc.} {\bfseries 258} (1992) 41}.

\bibitem{2004RvMP...76.1143P}
T.~{Piran}, \emph{{The physics of gamma-ray bursts}}, \href{https://doi.org/10.1103/RevModPhys.76.1143}{\emph{Reviews of Modern Physics} {\bfseries 76} (2004) 1143} [\href{https://arxiv.org/abs/astro-ph/0405503}{{\ttfamily astro-ph/0405503}}].

\bibitem{2006RPPh...69.2259M}
P.~{M{\'e}sz{\'a}ros}, \emph{{Gamma-ray bursts}}, \href{https://doi.org/10.1088/0034-4885/69/8/R01}{\emph{Reports on Progress in Physics} {\bfseries 69} (2006) 2259} [\href{https://arxiv.org/abs/astro-ph/0605208}{{\ttfamily astro-ph/0605208}}].

\bibitem{2014ARA&A..52...43B}
E.~{Berger}, \emph{{Short-Duration Gamma-Ray Bursts}}, \href{https://doi.org/10.1146/annurev-astro-081913-035926}{\emph{Annual Review of Astron and Astrophys} {\bfseries 52} (2014) 43} [\href{https://arxiv.org/abs/1311.2603}{{\ttfamily 1311.2603}}].

\bibitem{2015JHEAp...7...73D}
P.~{D'Avanzo}, \emph{{Short gamma-ray bursts: A review}}, \href{https://doi.org/10.1016/j.jheap.2015.07.002}{\emph{Journal of High Energy Astrophysics} {\bfseries 7} (2015) 73}.

\bibitem{2015PhR...561....1K}
P.~{Kumar} and B.~{Zhang}, \emph{{The physics of gamma-ray bursts \& relativistic jets}}, \href{https://doi.org/10.1016/j.physrep.2014.09.008}{\emph{Physics Reports} {\bfseries 561} (2015) 1} [\href{https://arxiv.org/abs/1410.0679}{{\ttfamily 1410.0679}}].

\bibitem{zhang_2018}
B.~Zhang, \emph{The Physics of Gamma-Ray Bursts}, Cambridge University Press (2018), \href{https://doi.org/10.1017/9781139226530}{10.1017/9781139226530}.

\bibitem{Ruffini2010}
R.~Ruffini, G.~Vereshchagin and S.-S.~Xue, \emph{Electron-positron pairs in physics and astrophysics: from heavy nuclei to black holes}, \href{https://doi.org/10.1016/j.physrep.2009.10.004}{\emph{Phys. Rept.} {\bfseries 487} (2010) 1} [\href{https://arxiv.org/abs/0910.0974}{{\ttfamily 0910.0974}}].

\bibitem{2012grb..book.....K}
C.~{Kouveliotou}, R.A.M.J.~{Wijers} and S.~{Woosley}, \emph{{Gamma-ray Bursts}}, Cambridge University Press (2012).

\bibitem{Ruffini1999}
R.~Ruffini, J.D.~Salmonson, J.R.~Wilson and S.-S.~Xue, \emph{On the pair electromagnetic pulse of a black hole with electromagnetic structure}, {\emph{Astron. Astrophys.} {\bfseries 350} (1999) 334} [\href{https://arxiv.org/abs/astro-ph/9907030}{{\ttfamily astro-ph/9907030}}].

\bibitem{Ruffini2000}
R.~Ruffini, J.D.~Salmonson, J.R.~Wilson and S.-S.~Xue, \emph{On the pair-electromagnetic pulse from an electromagnetic black hole surrounded by a baryonic remnant}, {\emph{Astron. Astrophys.} {\bfseries 359} (2000) 855} [\href{https://arxiv.org/abs/astro-ph/0004257}{{\ttfamily astro-ph/0004257}}].

\bibitem{Ruffini2003}
R.~Ruffini, L.~Vitagliano and S.-S.~Xue, \emph{On a separatrix in the gravitational collapse to an overcritical electromagnetic black hole}, \href{https://doi.org/10.1016/j.physletb.2003.08.051}{\emph{Phys. Lett. B} {\bfseries 573} (2003) 33} [\href{https://arxiv.org/abs/astro-ph/0309022}{{\ttfamily astro-ph/0309022}}].

\bibitem{Xue_2021}
S.-S.~Xue, \emph{{Gravo-thermal catastrophe in gravitational collapse and energy progenitor of Gamma-Ray Bursts}}, \href{https://doi.org/10.1088/1475-7516/2021/07/044}{\emph{JCAP} {\bfseries 07} (2021) 044} [\href{https://arxiv.org/abs/2104.03021}{{\ttfamily 2104.03021}}].

\bibitem{Xue2025}
S.-S.~Xue, \emph{Collimated and spinning fireballs for ultra-relativistic jets: long vs short gamma-ray bursts by angular momentum and mass ratio}, \href{https://doi.org/10.1140/epjc/s10052-025-14547-6}{\emph{Eur. Phys. J. C} {\bfseries 85} (2025) 820} [\href{https://arxiv.org/abs/2406.00454}{{\ttfamily 2406.00454}}].

\bibitem{Lemaitre:1933gd}
G.~Lemaitre, \emph{{The expanding universe}}, \href{https://doi.org/10.1023/A:1018855621348}{\emph{Annales Soc. Sci. Bruxelles A} {\bfseries 53} (1933) 51}.

\bibitem{Lasky:2006mg}
P.D.~Lasky and A.W.C.~Lun, \emph{{Generalized Lemaitre-Tolman-Bondi Solutions with Pressure}}, \href{https://doi.org/10.1103/PhysRevD.74.084013}{\emph{Phys. Rev. D} {\bfseries 74} (2006) 084013} [\href{https://arxiv.org/abs/gr-qc/0606055}{{\ttfamily gr-qc/0606055}}].

\bibitem{ODell1981}
S.L.~O'Dell, \emph{Radiation force on a relativistic plasma and the Eddington limit}, \href{https://doi.org/10.1086/183462}{\emph{\apjl} {\bfseries 243} (1981) L147}.

\bibitem{Sikora1981}
M.~Sikora, \emph{Superluminous accretion discs}, \href{https://doi.org/10.1093/mnras/196.2.257}{\emph{\mnras} {\bfseries 196} (1981) 257}.

\bibitem{Phinney1982}
E.S.~Phinney, \emph{Acceleration of a relativistic plasma by radiation pressure}, \href{https://doi.org/10.1093/mnras/198.4.1109}{\emph{\mnras} {\bfseries 198} (1982) 1109}.

\bibitem{Frederiksen_2008}
J.T.~Frederiksen, \emph{Stochastically induced gamma-ray burst wakefield processes}, \href{https://doi.org/10.1086/589648}{\emph{The Astrophysical Journal} {\bfseries 680} (2008) L5}.

\bibitem{DelGaudio2020}
F.~Del~Gaudio, R.A.~Fonseca, L.O.~Silva and T.~Grismayer, \emph{Plasma wakes driven by photon bursts via compton scattering}, \href{https://doi.org/10.1103/PhysRevLett.125.265001}{\emph{Phys. Rev. Lett.} {\bfseries 125} (2020) 265001} [\href{https://arxiv.org/abs/2003.04249}{{\ttfamily 2003.04249}}].

\bibitem{Martinez2021}
B.~Martinez, T.~Grismayer and L.O.~Silva, \emph{Compton-driven beam formation and magnetization via plasma microinstabilities}, \href{https://doi.org/10.1017/S0022377821000660}{\emph{J. Plasma Phys.} {\bfseries 87} (2021) 905870313} [\href{https://arxiv.org/abs/2102.11590}{{\ttfamily 2102.11590}}].

\bibitem{Faure2024}
J.C.~Faure, D.~Tordeux, L.~Gremillet and M.~Lemoine, \emph{High-energy acceleration phenomena in extreme-radiation{\textendash}plasma interactions}, \href{https://doi.org/10.1103/PhysRevE.109.015203}{\emph{Phys. Rev. E} {\bfseries 109} (2024) 015203} [\href{https://arxiv.org/abs/2309.00366}{{\ttfamily 2309.00366}}].

\bibitem{2014PhRvD..90a3009W}
Y.-B.~{Wu} and S.-S.~{Xue}, \emph{{Nonlinear Breit-Wheeler process in the collision of a photon with two plane waves}}, \href{https://doi.org/10.1103/PhysRevD.90.013009}{\emph{\prd} {\bfseries 90} (2014) 013009} [\href{https://arxiv.org/abs/1403.4798}{{\ttfamily 1403.4798}}].

\bibitem{Zhang:2018jje}
B.~Zhang, Z.-m.~Zhang, Z.-g.~Deng, W.~Hong, J.~Teng, S.-k.~He et~al., \emph{{Effects of Involved Laser Photons on Radiation and Electron-Positron Pair Production in one Coherence Interval in Ultra Intense Lasers}}, \href{https://doi.org/10.1038/s41598-018-35312-8}{\emph{Sci. Rep.} {\bfseries 8} (2018) 16862}.

\bibitem{Zhang2019}
B.~Zhang, Z.~Zhang, Z.-g.~Deng, J.~Teng, S.-k.~He, W.~Hong et~al., \emph{Quantum mechanisms of electron and positron acceleration through nonlinear compton scatterings and nonlinear breit-wheeler processes in coherent photon dominated regime}, \href{https://doi.org/10.1038/s41598-019-55472-5}{\emph{Scientific Reports} {\bfseries 9} (2019) 18876}.

\bibitem{2013CoPhC.184.2503X}
X.~{Xu}, P.~{Yu}, S.F.~{Martins}, F.S.~{Tsung}, V.K.~{Decyk}, J.~{Vieira} et~al., \emph{{Numerical instability due to relativistic plasma drift in EM-PIC simulations}}, \href{https://doi.org/10.1016/j.cpc.2013.07.003}{\emph{Computer Physics Communications} {\bfseries 184} (2013) 2503} [\href{https://arxiv.org/abs/1211.0953}{{\ttfamily 1211.0953}}].

\bibitem{1974ApJ...188..121B}
G.R.~{Blumenthal}, \emph{{The Poynting-Robertson Effect and Eddington Limit for Electrons Scattering with Hard Photons}}, \href{https://doi.org/10.1086/152693}{\emph{\apj} {\bfseries 188} (1974) 121}.

\bibitem{Exp_Sarri2025}
G.~Sarri, B.~King, T.~Blackburn, A.~Ilderton, S.~Boogert, S.S.~Bulanov et~al., \emph{Input to the European strategy for particle physics: strong-field quantum electrodynamics}, \href{https://doi.org/10.1140/epjp/s13360-025-07057-7}{\emph{The European Physical Journal Plus} {\bfseries 140} (2025) 11, 1151}.

\bibitem{böhm1989introduction}
E.~B{\"o}hm-Vitense, \emph{Introduction to Stellar Astrophysics: Volume 3}, Introduction to Stellar Astrophysics, Cambridge University Press (1989).

\end{thebibliography}

%\end{document}

\providecommand{\href}[2]{#2}\begingroup\raggedright\endgroup

\end{document}